\documentclass[journal,draftcls,onecolumn,12pt,twoside]{IEEEtran}

\usepackage{cite}

\ifCLASSINFOpdf
 
\else
  
\fi

\usepackage{algorithm}
\usepackage{algpseudocode}
\usepackage{array}

\usepackage{mdwmath}

\usepackage[tight,footnotesize]{subfigure}

\usepackage[font=footnotesize]{subfig}

\usepackage{tikz}
\usetikzlibrary{matrix,decorations.pathreplacing}
\usetikzlibrary{shapes}
\usetikzlibrary{plotmarks}
\usepackage{pgfplots}
\pgfplotsset{compat=1.12}
\usepgfplotslibrary{groupplots}
\usetikzlibrary{spy}
 \usepackage{pstricks-add}
 \usepackage{epsfig}
 \usepackage{tikz}
\usetikzlibrary{matrix,decorations.pathreplacing}
\usetikzlibrary{shapes}
\usetikzlibrary{plotmarks}
\usepackage{pgfplots}
\usetikzlibrary{spy}
\usepackage{pstricks-add}
\usepackage{epsfig}
\usepackage{pst-grad} 
\usepackage{pst-plot} 
\usepackage{collectbox}
\definecolor{darkblue}{rgb}{0.0, 0.0, 0.55}
\usepackage{tikz}
\usepackage{pgfplots}
\definecolor{darkorange}{rgb}{1.0, 0.55, 0.0}
\usepackage{graphicx}
\usepackage{epstopdf}
\usepackage{graphicx}
\usepackage{mathtools,textcomp}
\usepackage[numbers,sort&compress]{natbib}
\pgfdeclarelayer{background}
\pgfdeclarelayer{foreground}
\pgfsetlayers{background,main,foreground}
\tikzstyle{int}=[draw, fill=white!20, minimum size=2em]
\tikzstyle{init} = [pin edge={to-,thin,black}]
\usepackage{tikz}
\usetikzlibrary{shapes}
\usepackage{comment}
\usetikzlibrary{plotmarks}
\usepackage{pgfplots}
\usepgfplotslibrary{groupplots}
\usetikzlibrary{spy}

\usepackage{soul,color}

\tikzset{new spy style/.style={spy scope={%
			magnification=2.8,
			size=1.25cm,
			connect spies,
			every spy on node/.style={
				rectangle,
				draw,
			},
			every spy in node/.style={
				draw,
				rectangle,
			}
		}
	}
}

   	\pgfplotscreateplotcyclelist{mycolorlist}{%
	black,densely dashed,every mark/.append style={fill=black!80!black},mark=o\\
	brown,every mark/.append style={fill=brown!80!black},mark=otimes\\%
	red,every mark/.append style={fill=red!80!black},mark=otimes\\%
	blue,every mark/.append style={fill=blue!80!black},mark=otimes\\%
 magenta,every mark/.append style={fill=blue!80!black},mark=otimes\\%
 red,densely dashed,every mark/.append style={fill=black!80!black},mark=o\\
 blue,densely dashed,every mark/.append style={fill=black!80!black},mark=o\\
 brown,densely dashed,every mark/.append style={fill=black!80!black},mark=o\\
}

\usepackage{fixltx2e}

\usepackage{stfloats}

\begin{document}

\title{IRS-Assisted OTFS:  Beamforming Design  and Signal Detection \vspace{-3mm}}
	\author{
		\IEEEauthorblockN{\small{ Sushmita Singh$^{1}$, Kuntal Deka$^{2}$,  Sanjeev Sharma$^{3}$, and  Neelakandan Rajamohan$^{1}$ } }\\
		$^1$Indian Institute of Technology Goa, $^2$Indian Institute of Technology Guwahati,    $^3$Indian Institute of Technology (BHU) Varanasi, India  \vspace{-.5in}}

\maketitle

\begin{abstract}
Intelligent reflecting surface (IRS) technology has become a crucial enabler for creating cost-effective, innovative, and adaptable wireless communication environments.  
This study investigates an IRS-assisted orthogonal time frequency space (OTFS) modulation that facilitates communication between users and the base station (BS). The user's attainable downlink rate can be boosted by collaboratively improving the reflection coefficient (RC) matrix at the IRS and beamforming matrix at the BS. Then, in the IRS-aided OTFS network, the problem of cooperative precoding at BS and IRS to improve the network throughput is framed. The precoding design problem is non-convex and highly complicated; an alternate optimization (AO) approach is proposed to solve this. Specifically,  an approach based on strongest tap maximization (STM) and fractional programming is proposed. It solves RC matrix (at IRS) and beamforming matrix (at BS) alternatively. Moreover, an efficient signal detector for IRS-aided OTFS communication systems using the alternating direction method of multipliers (ADMM) is proposed. Finally,  to estimate the cascaded MIMO channel, using a parallel factor tensor model  that separates the IRS-User and BS-IRS MIMO channels, respectively is suggested. 
Simulation results show that the proposed method significantly enhances the system capacity and bit error rate (BER) performance compared to conventional OTFS.
\end{abstract}


\IEEEpeerreviewmaketitle

\vspace{-4mm}
\section{Introduction}  \label{sec1}
The 5G  and beyond wireless communication network is anticipated to have 1000 times more capacity than the 4G network \cite{intr2}. 
 To meet the demands for high-quality customer service, several vital advancements have been developed over the last decade. 
Recently, there have been significant advancements in intelligent reflecting surfaces (IRSs) and similar alternatives leveraging metamaterials. These developments hold great promise in delivering substantial improvements akin to multiple input multiple output (MIMO) systems, which were previously unimaginable. IRS acts as a planar array that can be dynamically reconfigured, comprising a multitude of passive reflecting elements.
By appropriately adjusting the attenuation and phase shifts of the constituent elements within the IRS, the reflected signals can be intelligently combined with other signal pathways at the receiver. This combination can be constructive, enhancing the received signal strength, or destructive, reducing co-channel congestion. In both cases, the result is an improved performance in wireless connections. These advantages make the IRS highly suitable for future wireless communications, offering affordability and sustainability.


Further,  a recently developed 2D modulation method called orthogonal time-frequency space (OTFS) can perform better than the widely used orthogonal frequency division multiplexing (OFDM) modulation in a variety of applications, including mmWave communications and vehicle to everything (V2X) communications \cite{hadaniotfs}.
 OTFS works in the delay-Doppler (DD) domain, where data symbols are placed in a 2D DD grid. By working in the DD domain, OTFS ensures that all symbols within a frame experience approximately similar channels.
This approach transforms the time-varying channel into an approximately time-invariant channel in the DD domain, providing consistent performance even in high Doppler conditions. 
Furthermore, OTFS achieves complete channel diversity through 2D localization in the DD domain, significantly enhancing overall performance.

Therefore, the combination of OTFS and IRS offers two key benefits:  improved received power gain and resistance against high Doppler. To fully leverage the advantages of OTFS enhanced by the IRS, optimizing both active (at BS) and passive (at IRS) precoding techniques is crucial. This optimization enables the extraction of the full benefits of OTFS in conjunction with IRS. Thus, this work considers the IRS-aided OTFS system by jointly optimizing beamforming and reflection coefficient (RC) matrix.


\subsection{Related Prior Works}
This section briefly overviews the previous works relevant to OTFS  and IRS technologies. A thorough investigation of various IRS-related topics was covered in the tutorial article \cite{irs1}.
 The effectiveness of IRS-aided systems using combined active and passive beamforming was examined in \cite{irs2}.
Various IRS-based scenarios have been explored in the literature to enhance the system's capacity. In particular, the authors in \cite{pa7} investigated a situation where a single BS and single IRS together serve a user, aiming to achieve capacity expansion using cost-effective and energy-efficient IRS. The case of multiple IRS scenarios was examined in \cite{pa01}.
The optimal positioning of the IRS   with respect to the BS and user's location and their ergodic capacity has been investigated in \cite{irs3}.  

The design and analysis of IRS-assisted systems have considered many existing technologies, such as OFDM. 
In \cite{ofdm_1}, a wireless system utilizing OFDM and IRS for frequency-selective channels was explored. The author in \cite{ofdm_2} maximized the downlink attainable sum rate of the user in an IRS-aided OFDM system by optimizing the transmit power distribution at the BS  and the RC phase shift at the IRS. 
The works in \cite{ofdm_5} proposed an effective channel estimation method for an IRS-aided OFDM system to reduce training time. The use of IRS in multichannel OFDM networks was studied in \cite{pa04} to increase the sum rate. In \cite{det_1}, the time-domain sparsity of the IRS-aided OFDM channel was exploited for joint channel estimation and data detection. 

Recently, IRS-aided OTFS has been investigated in \cite{thomas}. An RC phase shift matrix optimization was adopted where the most powerful DD link was co-phased with the direct link. Moreover, in \cite{thomas},  a channel estimation technique has been proposed for an IRS-aided OTFS network with little guard-band overhead where pilots and data were embedded in the same OTFS block.  In \cite{irs_stm}, the authors analyzed an IRS-aided MIMO OTFS communications system and proposed a minimum mean square error (MMSE) detection. For both single input single output (SISO) and MIMO scenarios, the IRS  RC phase shifts were designed to elevate the  SNR at the receiver side.

 
 
 \subsection{Contributions}
This paper focuses on the joint optimization of active beamforming (at BS) and RC (at IRS) for IRS-assisted OTFS systems. It ensures that the IRS contributes to increasing received power. The proposed precoding in the IRS-based OTFS network involves the architecture of the beamforming matrix at the BS and the RC phase adjustments at the IRS elements. Additionally, we develop  signal detection method of IRS-aided OTFS systems. The key contributions in this paper are highlighted below:
 \begin{enumerate}
     \item  First, we derive the input-output relation for the SISO IRS-aided  OTFS system, and then we extend the derivation to the MIMO case. After that, we address the challenge of joint design of the beamforming matrix at the BS and the RC matrix at the IRS to maximize the average sum rate (ASR) for the  user. This joint optimization increases the network capacity for the envisioned IRS-aided OTFS network.
     \item 
    We propose an alternating optimization (AO) framework. We first optimize the RC at the IRS with the strongest tap maximization (STM) method. Later, we use multidimensional complex quadratic transformation (MCQT) and Lagrange dual reformulation (LDR) to optimize the beamforming matrix in the realms of fractional programming. 
     \item Furthermore, we validate the efficacy of the proposed joint beamforming matrix and RC phase shift optimization framework by presenting simulation results that support our theoretical conclusions on the IRS-aided OTFS network.
 Results from simulations show that IRS can significantly increase the capacity of networks. In addition, it is essential to note that, due to the applicability of the investigated topic, the suggested AO algorithm can also be used as a generic solution to increase the sum rate in the majority of the  IRS-aided scenarios explored so far in the literature.
\item To the authors' knowledge, this work is the first attempt to detect the OTFS symbols using the alternating direction method of
multipliers (ADMM) algorithm. 
A comprehensive performance analysis of the ADMM algorithm is conducted  for detecting IRS-aided OTFS signals. We opted for the ADMM detector over message passing (MP) and MMSE detectors. Notably, the channel cascading effect in an IRS-assisted OTFS system diminishes the sparsity within the underlying DD-domain, rendering MP impractical for detection. Furthermore, the performance of the MMSE detector notably deteriorates in comparison to the proposed ADMM-based detector.
\item  Lastly, an effective method  is proposed to estimate the cascaded MIMO channel using a parallel factor tensor model. This method adopts an iterative alternating least square (ALS) algorithm for the channel estimation.

 \end{enumerate}
 \subsection{Organization of the Paper}
The paper is organized as follows:
In Section~\ref{sec2}, we analyze the mathematical model of the IRS-aided OTFS and the essential preliminaries of fractional programming, and the ADMM algorithm. Section~\ref{sec3} provides the problem formulation of the ASR maximization. Joint active and passive precoding architecture is suggested to solve the formulated problem.  
Further, Section~\ref{sec4} and Section~\ref{chaest} cover the ADMM-based detection approach and the channel estimation algorithm for the IRS-OTFS system, respectively. Section~\ref{sec5} includes simulation results. Finally, Section~\ref{sec6} concludes the paper.

\noindent \textit{Notations:}
Every lowercase letter $x$ used in this paper represents a scalar. A vector is represented by the bold lowercase letter $\mathbf{x}$, while a matrix of the given dimension is represented by the bold uppercase letter $\mathbf{X}$. ${\mathbf{X}}^{H}$, ${\mathbf{X}}^{T}$, $\textbf{X}^{\dagger}$ represents conjugate transpose, transpose, the pseudo-inverse and the phase of any complex number $a$ is shown by the symbol $\angle a$. The $N$-point FFT matrix is represented by $\textbf{F}_{N}$. The set of all real numbers and the set of all complex numbers are denoted by $\mathcal{R}$ and $\mathcal{C}$, respectively. ${\text{Re}}(\cdot)$ denotes the real part.  $\textbf{I}_{N}$ represents the identity matrix of dimension $N \times N$. $\text{vec}(\cdot)$ represents the vectorization of the matrix. The convolution operation and Kronecker product are denoted by $*$ and $\otimes $, respectively.

\section{System Model of the Proposed IRS-assisted OTFS System}  \label{sec2}
\subsection{\label{a}Preliminaries}
In OTFS, a 2D modulation method is employed, consisting of a DD grid with dimensions $M \times N$, where $M$ and $N$ represent the number of delay and Doppler bins, respectively. The overall bandwidth for an OTFS frame at a particular sub-carrier frequency  $\Delta f$, with $T \Delta f =1$, is $B = M \Delta f $, while the total frame transmission duration is given by $T_{f} = NT $. The parameters $M$ and $N$ are chosen to ensure sufficient delay resolution $T = \frac{1}{M\Delta f}$ and Doppler resolution $\nu = \frac{1}{NT}$.

\subsubsection{Basics of OTFS Transceiver}
Information symbols (like quadrature amplitude modulation (QAM) symbols) are placed over the DD domain in the OTFS modulation. The transmitter uses the inverse symplectic finite Fourier transform (ISFFT) to translate the DD domain symbol $\textbf{A}^{\mathrm{DD}}[l,k]$ into the frequency-time (FT) domain symbol $\textbf{A}^{\text{FT}}[m,n]$ and is given as $\textbf{A}^{\mathrm{FT}}[m,n] = \frac{1}{\sqrt{MN}}\sum_{k=0}^{N-1}\sum_{l=0}^{M-1}\textbf{A}^{\mathrm{DD}}[l,k]e^{-j2\pi(\frac{nl}{M}-\frac{nk}{N})}$. Here, the number of Doppler and delay bins in the DD domain, respectively, are $N$ and $M$. $\textbf{A}^{\mathrm{FT}}[m,n]$ is transformed using the Heisenberg transform to produce a time-domain transmission signal $s(t)$ as $s(t) = \sum_{n=0}^{N-1}\sum_{m=0}^{M-1}\textbf{A}^{\mathrm{FT}}[m,n]e^{j2\pi m\Delta f(t-nT)}p_{\text{tx}}(t-nT)$, where $p_{\text{tx}}(t)$ denotes the transmit pulse shape. The received signal $r(t)$ in the time-domain   is given by $r(t) = \int_{}^{}\int_{}^{}h(\tau, \nu)e^{j 2\pi \nu(t-\tau)}s(t-\tau)d\tau d\nu$, where the channel impulse response $h(\tau,\nu)$ is defined by the the Doppler frequency $\nu$ and delay $\tau$ and $h(\tau,\nu) = \sum_{l=0}^{L}h_{l}e^{j\theta_{l}}\delta(\tau - \tau_{l})\delta(\nu -\nu_{l})$. Here, $\delta(\cdot)$ stands for the delta function, $L$ is the total number of paths. Further,  $h_{l}$, $\tau_{l}$, and $\nu_{l}$ are the channel coefficient, delay, and Doppler frequency respectively for the $l$-th  path.   The Doppler bins and the delay are represented by the integer indices
$k_{\nu_{l}} $ and $l_{\tau_{l}}$ and the fractional Doppler is represented by $\kappa_{\nu_{l}}$  such that $\tau_{l} = \frac{l_{\tau_{l}}}{M\Delta f}$ and $\nu_{l} = \frac{k_{\nu_{l}}+\kappa_{\nu_{l}}}{NT}$.
 The receiver conducts matched filtering utilizing the pulse shape filter $p_{\text{rx}}(t)$ to obtain $\textbf{B}_{p_{\text{rx}},r}(\tau,\nu) \overset{\Delta}{=} \int_{}^{}e^{-j 2 \pi \nu(t-\tau)}p^{*}_{\text{rx}}(t-\tau)r(t)dt$. Now the FT domain received signal is obtained as: $\textbf{Y}^{\text{FT}}[m,n]= \textbf{B}_{P_{\text{rx}},r}(\tau,\nu)|_{\tau=nT,\nu = m\Delta f}$. The receiver then uses the symplectic finite Fourier transform (SFFT) to retrieve the received symbols in the DD domain and is given by  $\textbf{Y}^{\text{DD}}[m,n] = \frac{1}{\sqrt{MN}}\sum_{n=0}^{N-1}\sum_{m=0}^{M-1}\textbf{Y}^{\text{FT}}[m,n]e^{-j 2\pi (\frac{nk}{N}-\frac{ml}{M})}$.

 The transmit and receive pulse shapes are rectangular. The transmit pulse is defined by $\frac{1}{\sqrt{T}}$ with $t\in [0,T)$ and $0$ otherwise. The receive pulse is given by $\frac{1}{\sqrt{T}}$ with $t\in [-T_{\mathrm{CP}},T)$ and $0$ otherwise, where $T_{\mathrm{CP}}$ is the length of CP. The transmitted signal $s(t)$ can be modeled as a symbol-by-symbol block $\textbf{S}\in \mathcal{C}^{M\times N}$ by introducing the rectangular transmit pulse form and is given by: $\textbf{S} = \frac{1}{\sqrt{T}}\textbf{F}_{N}^{H}\textbf{A}^{\mathrm{FT}} = \frac{1}{\sqrt{T}}\textbf{A}^{\mathrm{DD}}\textbf{F}_{N}^{H}$, where $\textbf{A}^{\mathrm{FT}}$, $\textbf{A}^{\mathrm{DD}} \in \mathcal{C}^{M\times N}$ and it comprises of $\textbf{A}^{\mathrm{FT}}[m,n]$ and $\textbf{A}^{\mathrm{DD}}[l,k]$ respectively. $\textbf{F}_{M}$ and $\textbf{F}_{N}$ are the $M$-point and $N$-point discrete Fourier transform (DFT) matrices respectively. Let $\textbf{G}_{\mathrm{CP}}\in \mathcal{C}^{(M+M_{\mathrm{CP}})\times M} $ be the CP matrix which is to be added to prevent interference within the OFDM symbols, where $M_{\text{CP}}$  is the CP length in terms of sub-carriers. Lastly, in order to produce a transmit signal containing CPs in the time domain, the transmitter undergoes parallel to serial conversion and is given by: $\textbf{s} = \text{vec}(\textbf{G}_{\mathrm{CP}}\textbf{S} ) = \frac{1}{\sqrt{T}}\text{vec}(\textbf{G}_{\mathrm{CP}} \textbf{A}^{\mathrm{DD}}\textbf{F}_{N}^{H})$. Following the removal of the CP, the $n$-{th} received OFDM symbol $\textbf{r}_{n} \in \mathcal{C}^{M}$ is given by: $\textbf{r}_{n} = \textbf{H}_{n}\textbf{s}_{n} + \textbf{z}_{n}$ where  $\textbf{z}_{n}$ is the additive white Gaussian noise and $\textbf{H}_{n} \in \mathcal{C}^{M \times M}$ is the $n$-th OFDM symbol's channel matrix, and can be written as
\begin{equation}\label{eq1}
    \textbf{H}_{n} = \sum_{q=1}^{L}h_{q}e^{j \theta_{q}}\boldsymbol{\Delta}_{n}^{k^{q}_{\nu},l^{q}_{\tau}}\boldsymbol{\Pi}^{l^{q}_{\tau}},
\end{equation}
where $\boldsymbol{\Pi}^{l_{\tau}^{q}} \in \mathcal{R}^{M \times M}$ and $\boldsymbol{\Delta}_{n}^{k_{\nu}^{q},l_{\tau}^{q}} \in \mathcal{C}^{M \times M}$ denotes the delay as well as Doppler shift matrices for the $q$-th path \cite{cpofdm1}. To acquire the symbol vector in the frequency domain, the receiver conducts $M$-point DFT in the time domain upon its $n$-~th OFDM symbol. Hence, $\textbf{y}_{n}^{\mathrm{FT}}= \textbf{F}_{M}\textbf{H}_{n}\textbf{A}^{\mathrm{DD}}\textbf{f}_{n}^{\ast }+\textbf{F}_{M}\textbf{z}_{n}$ where $\textbf{f}^{*}_{n} \in \mathcal{C}^{N}$ is the $n$-th column vector of $\textbf{F}_{n}$. Suppose $\textbf{Y}^{\mathrm{FT}}\overset{\Delta}{=}\left[ \textbf{x}_{1}^{\mathrm{FT}},\textbf{x}_{2}^{\mathrm{FT}},...,\textbf{x}_{N}^{\mathrm{FT}} \right] \in \mathcal{C}^{M \times N}$ is the FT domain two dimensional received signal. After that, the  SFFT operation is applied on $\textbf{Y}^{\text{FT}}$. The  symbol matrix in the DD domain can be obtained as $\textbf{Y}^{\text{DD}} = \textbf{F}_{N}^{H}\textbf{Y}^{\text{FT}}\textbf{F}_{N}$. $\textbf{Y}^{\text{DD}}$ can be written in vectorised form as  $\textbf{y}^{\mathrm{DD}}  = \text{vec}\left(\textbf{Y}^{\mathrm{DD}}  \right) =\left( \textbf{F}_{N}\otimes \textbf{I}_{M} \right)\textbf{H}_{\text{eff}}\left(\textbf{F}_{N}^{H}\otimes \textbf{I}_{M}  \right)\textbf{a}^{\mathrm{DD}} +\left( \textbf{F}_{N}\otimes \textbf{I}_{M} \right)\textbf{z}$. Thus, the input-output relation is expressed as
\begin{equation}
\textbf{y}^{\mathrm{DD}}= \tilde{\textbf{H}}_{\text{eff}}\textbf{a}^{\mathrm{DD}}+\left( \textbf{F}_{N}\otimes \textbf{I}_{M} \right)\textbf{z},
\end{equation}
where
$\tilde{\textbf{H}}_{\text{eff}} = \left( \textbf{F}_{N}\otimes 
\textbf{I}_{M} \right)\textbf{H}_{\text{eff}}\left(\textbf{F}_{N}^{H}\otimes \textbf{I}_{M}  \right)$, $\textbf{a}^{\mathrm{DD}} = \text{vec}\left(\textbf{A}^{\mathrm{DD}}  \right)$, and 
$\textbf{H}_{\text{eff}} = \text{diag}\left[\textbf{H}_{1},\textbf{H}_{2},...,\textbf{H}_{N}  \right]
$.


\subsection{MIMO IRS-aided  OTFS System }
\begin {figure}[!htbp]
\centering
	\includegraphics[width=4.5in, height=3in]{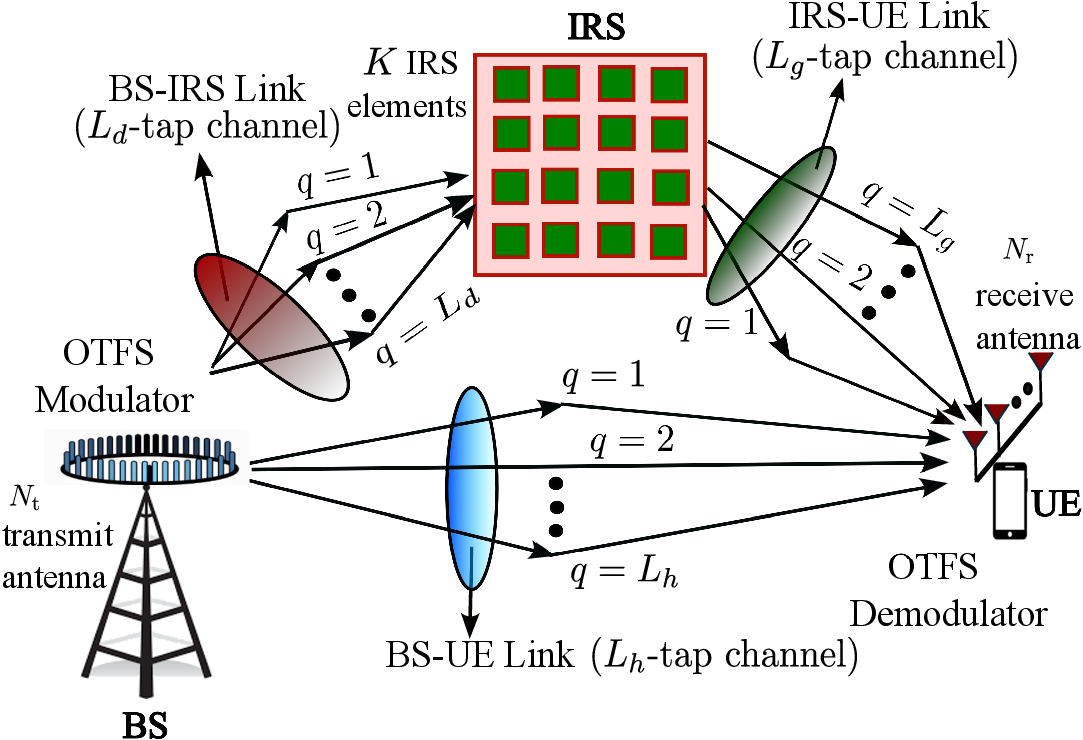}\\
	\caption{\footnotesize{ IRS-aided MIMO OTFS system.}}
	\label{fig1}
 \vspace{-6mm}
\end{figure}

 Fig.~\ref{fig1} depicts an IRS-OTFS MIMO system, where the BS involves $N_{\mathrm t}$ transmit antennas, and these signals are reflected by $K$ IRS elements before reaching $N_{\mathrm r}$ receive antennas. 
It is assumed that
 the channels between BS and IRS, IRS and UE, and BS to UE  exhibit time-varying characteristics. 
Note that, typically, in the DD domain, the channel has a sparse representation.
The beamforming matrix for the $j$-th subcarrier of the $n$-th OFDM block is denoted as $\textbf{V}_{j} \in \mathcal{C}^{N_{\mathrm t} \times N_{\mathrm r}}$. Consequently, we can express the beamforming matrix for the $n$-th OFDM symbol of the OTFS block as follows \cite{W}:
\begin{equation}
\label{115}
    \textbf{W}_{n} = \begin{bmatrix}
\textbf{V}_{0} &  &  \\
 &\ddots  &  \\
 &  &\textbf{V}_{M-1} 
\end{bmatrix}\in \mathcal{C}^{MN_{\mathrm t} \times MN_{\mathrm r}}.
\end{equation}


  The  time-domain input-output relationship  is expressed as follows   
\begin{equation}\label{h1}
 \textbf{r}_{n} = \left(\textbf{H}^{d}_{n} + \textbf{D}_{n}\boldsymbol{\Theta}_{n}\textbf{G}_{n}\right)\textbf{W}_{n}\textbf{s}_{n}+\textbf{z}_{n} .  
\end{equation}
Initially, $\textbf{W}_{n}$ digitally precodes the vector $\textbf{s}_{n}$ in the frequency domain before transmitting it and then we  convert the signal to the time domain as
\begin{equation}
    \tilde{\textbf{r}}_{n} = (\textbf{H}^{d}_{n} + \textbf{D}_{n}\boldsymbol{\Theta}_{n}\textbf{G}_{n})\tilde{\textbf{s}}_{n}+\textbf{z}_{n},
    \label{received}
\end{equation}
where  $\tilde{\textbf{s}}_{n} = \textbf{W}_{n}\textbf{s}_{n}$ and $\textbf{s}_{n} = \left[ \textbf{s}_{n,1}^{T},\textbf{s}_{n,2}^{T},....,\textbf{s}_{n,N_{\mathrm t}}^{T} \right]^{T}\in \mathcal{C}^{N_{\mathrm t}M}$ is the transmitted time domain signal of $n$-th OFDM symbol of OTFS frame and $\textbf{r}_{n} = \left[ \textbf{r}_{n,{1}}^{T},\textbf{r}_{n,{2}}^{T},....,\textbf{r}_{n,{N_{\mathrm r}}}^{T} \right]^{T}\in \mathcal{C}^{N_{\mathrm r}M}$ is the respective  received signal. In (\ref{received}), $\boldsymbol{\Theta}_{n}$ is a diagonal RC matrix of the $n$-th OFDM symbol of OTFS frame   comprising of $K$ IRS elements, i.e., 
\begin{equation}\label{theta}
\boldsymbol{\Theta}_{n}={\text{diag}}(\beta_{1}e^{j\theta_{1}},\beta_{2}e^{j\theta_{2}},....,\beta_{k}e^{j\theta_{K}})\otimes \textbf{I}_{M}.
\end{equation}
$\textbf{H}^{d}_{n}$, $\textbf{G}_{n}$, and $\textbf{D}_{n}$  are the channel matrices for BS to UE also known as direct link (D-link), BS to IRS, and IRS to UE channels, respectively for $n$-th OFDM symbol of IRS-assisted OTFS system.
The structures of these matrices are explained below:
\begin{align}
    \textbf{H}^{d}_{n} = \left\{ \textbf{H}^{d}_{n,n_{\mathrm r},n_{\mathrm t}} \right\}_{n_{\mathrm t}=1,n_{r}=1}^{N_{\mathrm t},N_{\mathrm r}}, \ 
    \textbf{G}_{n} = \left\{ \textbf{G}_{n,k,n_{\mathrm t}} \right\}_{k=1,n_{\mathrm t}=1}^{K,N_{\mathrm t}}, \  
    \textbf{D}_{n} = \left\{ \textbf{D}_{n,n_{\mathrm r},{k}} \right\}_{n_{\mathrm r}=1,k=1}^{N_{\mathrm r},K}
\end{align}
where $\textbf{H}^{d}_{n,n_{\mathrm t},n_{\mathrm r}}, \textbf{G}_{n,k,n_{\mathrm t}},\textbf{D}_{n,n_{\mathrm r},k}\in \mathcal{C}^{M \times M}$ are  the channels between $n_{{\text{r}}}$-th receive antenna and the $n_{\text{t}}$-th transmit antenna, between the $n_{\text{t}}$-th transmit antenna and the $k$-th IRS element, and between the $k$-th IRS element and the $n_{{\text{r}}}$-th receive antenna, respectively of $n$-th OFDM symbol of IRS-aided OTFS signal.
Now the cascaded channel matrix for the $n$-th OFDM symbol following (\ref{h1}) can be written as  
\begin{equation*}
    \textbf{H}_{n} = \textbf{H}^{d}_{n}+(\textbf{D}_{n}\boldsymbol{\Theta}_{n}\textbf{G}_{n}) \ \text{and} \  \textbf{H}_{b_{n}} = \textbf{H}^{d}_{n}+(\textbf{D}_{n}\boldsymbol{\Theta}_{n}\textbf{G}_{n})\textbf{W}_{n}.
\end{equation*}
Here, $\textbf{H}_{b_{n}}$ is the cascaded channel matrix of the $n$-th OFDM symbol along with the precoding matrix.
Now the cascaded channel matrix of IRS aided MIMO-OTFS system can be written as
\begin{equation}\label{H4}
    \textbf{H}_{b} = {\text{diag}}(\textbf{H}_{b_{1}},\textbf{H}_{b_{2}},....,\textbf{H}_{b_{N}}).
\end{equation}
The objective is to optimize the capacity of the beamformed  IRS-aided MIMO OTFS system. The expression of the capacity is given in Section~\ref{sec3} and to maximize the capacity, the tools of fractional programming (FP) are used. The next subsection provides the preliminaries of fractional programming.

\subsection{Fractional Programming (FP)}
FP is used to optimize the ratio of two functions subject to certain constraints. 
One primary interest in communication systems is to optimize the data rate, i.e., $\log(1+\text{SNR})$, where SNR denotes the signal-to-noise ratio. 
FP techniques involve two key transformations, i.e., Lagrangian dual transformation and quadratic transformation to simplify the problem.  These transformations are discussed below.

\subsubsection{Lagrange's Dual Transformation}
The Lagrangian dual transform method is effective in moving SNR outside of the logarithmic term $\log(1+\text{SNR})$. This subsection provides a thorough explanation of the Lagrangian dual transform method.
\begin{itemize}
    \item $Lemma\hspace{2mm} 1:$ \newline
    Consider a series of complex, multidimensional functions ${\mathbf{P(\textbf{x})}}$ and   ${\mathbf{Q(\textbf{x})}}$.  Suppose $\mathcal{X}$ is a collection of nonempty constraints. Then consider a  multidimensional and complex logarithmic FP problem  given by
    \begin{align} \label{fp1}
        \underset{\textbf{x}}{\max}\hspace{3mm}\sum_{m=1}^{M}w_{m}\hspace{1.5mm}\log\left(1+\textbf{P}_{m}^{H}(\textbf{x})\textbf{Q}_{m}^{-1}(\textbf{x})\textbf{P}_{m}(\textbf{x})\right) \quad \text{s.t.} \hspace{5mm}\textbf{x}\in \mathcal{X},
    \end{align}
    where, $w_{m}$ is a non negative weight with $m = 1,2,....,M$. 
    
    The problem (\ref{fp1})  is equivalent to
    \begin{align}
    \label{fp_modified}
        \underset{\textbf{x},\boldsymbol{\gamma}}{\max}\hspace{3mm}R_{r}(\textbf{x},\boldsymbol{\gamma}) \quad 
        \text{s.t.} \hspace{5mm} \textbf{x}\in \mathcal{X},
    \end{align}
    where $\boldsymbol{\gamma}= \left[\gamma_{1},....,\gamma_{M} \right]^T$ is a set of new auxiliary variables that are added for every ratio term. The newly introduced objective function $R_{r}$ is described by \cite{fp2}:
    \begin{align}\label{fp2}
        &R_{r}(\textbf{x},\boldsymbol{\gamma}) = \sum_{m=1}^{M}w_{m}\hspace{1.5mm}\log(1+\gamma_{m}) - \sum_{m=1}^{M}w_{m}\gamma_{m} +  \nonumber  \\
        & \sum_{m=1}^{M}w_{m}(1+\gamma_{m})\textbf{P}_{m}^{H}(\textbf{x})(\textbf{P}_{m}(\textbf{x})\textbf{P}_{m}^{H}(\textbf{x})+\textbf{Q}_{m}(\textbf{x}))^{-1}\textbf{P}_{m}(\textbf{x}).
    \end{align}
    It is proved in \cite{fp2}  that the problems in   (\ref{fp1})  and (\ref{fp_modified}) are equivalent paving the way for a simplified procedure of optimization. 
\end{itemize}

\subsubsection{Quadratic Transform}
The quadratic transform converts the ratios into quadratic form in an effort to simplify the optimization task. The following Lemma shows the equivalence of the solutions post quadratic transformation.
\begin{itemize}
    \item $Lemma\hspace{2mm}2:$\newline
      Consider a series of complex, multidimensional functions ${\mathbf{P(\textbf{x})}}$ and   ${\mathbf{Q(\textbf{x})}}$.  Suppose $\mathcal{X}$ is a collection of nonempty constraints. Then consider a  multidimensional and complex logarithmic FP problem as given by \cite{fp2}
     \begin{align}\label{fp3}
         \underset{\textbf{x}}{\max}\hspace{3mm}\sum_{m=1}^{M}\textbf{P}_{m}^{H}(\textbf{x})(\textbf{Q}_{m}(\textbf{x}))^{-1}\textbf{P}_{m}(\textbf{x})  \quad  \text{s.t.}\hspace{5mm}\textbf{x}\in \mathcal{X}.
     \end{align}
    The above problem in (\ref{fp3}) is equivalent to the following problem \cite{fp2}:
     \begin{align}\label{fp4}
         \underset{\textbf{x},\boldsymbol{\alpha}}{\max}\hspace{3mm}\sum_{m=1}^{M}2\hspace{1mm}\mathcal{R}\left\{ \alpha_{m}^{H}\textbf{P}_{m}(\textbf{x}) \right\}-\alpha_{m}^{H}\textbf{Q}_{m}(\textbf{x})\alpha_{m} \quad 
          \text{s.t.}\hspace{5mm}\textbf{x}\in \mathcal{X},
     \end{align}
     where $\boldsymbol{\alpha}$ refers to $(\alpha_{1},\alpha_{2},....,\alpha_{M})$.
\end{itemize}

\subsection{ADMM Algorithm}
ADMM is an optimization algorithm that is often used to solve convex optimization problems.
ADMM is particularly useful when dealing with problems that have certain structural properties, such as separable objective functions or constraints. By segmenting the fundamental convex problem into acceptable sub-problems, ADMM can solve it efficiently. ADMM combines the dual ascent's decomposability with the method
of multipliers's improved convergence characteristics. The algorithm addresses problems of the following form
 \begin{align}
    &\min\hspace{5mm}f(\textbf{x})+g(\textbf{y})  \nonumber  \\
    &\text{s.t.}\hspace{3mm}\text{A}\textbf{x}+\text{B}\textbf{y}  = \textbf{z}.
\end{align}
where $f(\cdot)$ and $g(\cdot)$ are assumed to be convex.
The augmented Lagrangian  is given by
\begin{align}
    L_{\rm P}(\textbf{x},\textbf{y},\textbf{s}) = f(\textbf{x})+g(\textbf{y})+\textbf{s}^{T}(\text{A}\textbf{x}+\text{B}\textbf{y}-\textbf{z})+ 
    \left(\frac{\rho}{2}\right)\left\| \text{A}\textbf{x}+\text{B}\textbf{y}-\textbf{z} \right\|_{2}^{2}.
\end{align}
ADMM consists of the following steps
\begin{align}
    & \textbf{x}^{i+1} = \underset{\textbf{x}}{\text{argmin}}\hspace{1mm}L_{\rm P}(\textbf{x},\textbf{y}^{i},\textbf{s}^{i})  \nonumber   \\
    & \textbf{y}^{i+1} = \underset{\textbf{y}}{\text{argmin}}\hspace{1mm}L_{\rm p}(\textbf{x}^{i+1},\textbf{y},\textbf{s}^{i})  \nonumber  \\
    & \textbf{s}^{i+1}= \textbf{s}^{i} + \rho(\text{A}\textbf{x}^{i+1}+\text{B}\textbf{y}^{i+1}-\textbf{z}), 
\end{align}
where $i$ is the iteration index and  $\rho$ ($\rho >0$) denotes the penalty parameter.  The approach is a combination of  the dual ascent technique and the multiplier method.

\section{Beamforming for IRS-aided MIMO-OTFS System} \label{sec3}
This section discusses the active as well as passive beamforming techniques for the IRS-aided MIMO-OTFS system. 
\subsection{Capacity Analysis}
An OTFS symbol is transmitted using $N$ successive transmissions of OFDM blocks. The transmission of $N$ parallel OFDM blocks can be treated equivalently to that of a single OTFS block. We assume the channel to be ergodic and independent for each subsequent OTFS block.
The mutual information between the transmit data vector $\textbf{a}^{\mathrm{DD}}$ and the IRS-aided OTFS received signal vector \textbf{r} is given by \cite{mimoofdmotfs}
\begin{equation}
I(\textbf{r};\textbf{a}^{\text{DD}}) = \sum_{n=0}^{N-1}\log_{2}\left( \frac{\left| \textbf{H}_{b_{n}}\textbf{H}_{b_{n}}^{H} +\sigma^{2}\textbf{I}_{MN_{\mathrm r}} \right|}{\left| \sigma^{2} \textbf{I}_{MN_{\mathrm r}}\right|} \right) \nonumber.
\end{equation}
 
Therefore, the capacity is given by
\begin{equation}
\label{24}
{R}_{\text{OTFS}} = \frac{1}{M+M_{\mathrm{CP}}}\sum_{n=0}^{N-1}\log_{2}\left( \frac{\left| \textbf{H}_{b_{n}}\textbf{H}_{b_{n}}^{H} +\sigma^{2}\textbf{I}_{M N_{\mathrm r}} \right|}{\sigma^{2^{MN_{\mathrm r}}}} \right).
\end{equation}
 
 \subsection{Problem  Formulation}
We alternately optimize the active beamforming matrix at the BS and the passive RC matrix at the IRS to maximize the attainable rate of the downlink IRS-enhanced MIMO-OTFS system. Assume that the maximum power available to the BS is $P_{\text{max}}$. Hence, we have transmit power constraint as: $\left\| \textbf{W} \right\|^{2}\le P_{\text{max}}$, where $\textbf{W} = \text{diag}\left(  \textbf{W}_{0},\textbf{W}_{1},...,\textbf{W}_{N-1} \right)$. The achievable rate can be expressed as
\begin{equation}\label{R}
    R(\textbf{W},\boldsymbol{\Theta)} = \frac{1}{M+M_{CP}}\sum_{n=0}^{N-1}\log_{2}\left( \frac{\left| \textbf{H}_{b_{n}}\textbf{H}_{b_{n}}^{H} +\sigma^{2}\textbf{I}_{MN_{r}} \right|}{\sigma^{2^{MN_{\mathrm r}}}} \right).
\end{equation}
To make the subsequent optimization problem more concise, we remove the  constant term in (\ref{R}), and the optimization problem becomes
\begin{align}
\label{27}
(P0):\hspace{2mm} &\underset{\textbf{W},\boldsymbol{\Theta}}{\max}\hspace{1.5mm}\sum_{n=0}^{N-1}\log_{2}\left( \frac{\left| \textbf{H}_{b_{n}}\textbf{H}_{b_{n}}^{H} +\sigma^{2}\textbf{I}_{MN_{r}} \right|}{\sigma^{2^{MN_{\mathrm r}}}} \right) \nonumber \\
& \text{s.t.} \hspace{2mm} \sum_{n=0}^{N-1} \left\| \textbf{W}_{n} \right\|^{2}\le P_{\max}  \nonumber  \\
&  \hspace{7mm} \theta_{k} \in (-\pi,\pi].
\end{align}
The problem aims to optimize the beamforming matrices \textbf{W} and the phase shifts $\boldsymbol{\Theta}$ in order to maximize the total capacity of the MIMO system, while taking into account actual constraints on power and phase shifts.
The joint optimization of the  RC  matrix $\boldsymbol{\Theta}$ and the beamforming matrix $\mathbf{W}$  is difficult due to the  non-convex objective function. Also, the objective function involves a complex relationship with the RC matrix and the beamforming matrix, making it even more challenging to solve. Here, we develop a low-complexity algorithm to solve the above objective function effectively. 

\subsection{\label{AO} Algorithm for Joint Beamforming Optimization}
In this subsection, we elaborate  the joint beamforming optimization algorithm by splitting the main optimization problem into a number of manageable sub-problems. The algorithm is explained below.
\subsubsection{Passive Precoding with Fixed Active Precoding}
To improve the received SNR, we adopt the STM method \cite{stm1} in this section for determining a collection of IRS phase shifts in the proposed OTFS system by considering a  fixed $\textbf{W}_{n}$ which is initially randomly generated. 
This technique chooses the IRS reflection to align the strongest cascaded link with the direct link. For  analysis, we assume unit transmit power. The problem formulation for determining the IRS RC can be expressed as:
\begin{equation}
    \underset{\boldsymbol{\Theta}_{n}}{\max}\left\| \left( \textbf{D}_{n}\boldsymbol{\Theta}_{n}\textbf{G}_{n} \right) \textbf{W}_{n}\right\|^{2}/\sigma^{2} \hspace{7mm}\text{s.t.} \hspace{4mm}\theta_{k}\in (-\pi,\pi].
\end{equation}
For simplicity, we drop $\textbf{W}_{n}$ as it has been made constant. 
The above objective function can be enlarged as
\begin{equation}\label{eq22}
\left\|
\begin{aligned}
     e^{j\theta_{k}}\sum_{\tilde{p}=1}^{L_{d}} \sum_{\hat{p}=1}^{L_{g}}d_{n_{\text{t}},k}^{(\tilde{p})}g_{k,n_{\text{r}}}^{(\hat{p})}&\boldsymbol{\Pi}^{l_{d_{n_{\text{t}},k}}^{(\tilde{p})}+l_{g_{k,n_{\text{r}}}}^{(\hat{p})}}   \\
&\boldsymbol{\Delta}_{n}^{k_{d_{n_{\text{t}},k}}^{(\tilde{p})}+k_{g_{k,n_{\text{r}}}}^{(\hat{p})},l_{d_{n_{\text{t}},k}}^{(\tilde{p})}+l_{g_{k,n_{\text{r}}}}^{(\hat{p})}}
\end{aligned}
\right\|^{2},
\end{equation}
where $\left( d_{n_{\text{t}},k}^{\tilde{p}},g_{k,n_{\text{r}}}^{\hat{p}} \right)$,$\left(l_{d_{n_{\text{t}}},k}^{\tilde{p}},l_{g_{k,n_{\text{r}}}}^{\hat{p}}\right)$ and $\left(k_{d_{n_{\text{t}}},k}^{\tilde{p}},k_{g_{k,n_{\text{r}}}}^{\hat{p}} \right)$ are the channel coefficient, delay,  and the Doppler shift, respectively,  for the $\tilde{p}$-${\text{th}}$ path of BS to IRS and $\hat{p}$-${\text{th}}$ path of IRS to UE. 
Finally, the optimization problem can be written as \cite{thomas}
\begin{equation}
    \underset{\tilde{p},\hat{p},\theta_{k}}{\max}\hspace{3mm}\sum_{n=0}^{N-1}\left| e^{j\theta_{k}}d_{n_{\text{t}},k}^{\tilde{p}}g_{k,n_{\text{r}}}^{\hat{p}} \right|^{2}.
\end{equation}
First,  observe that for a given path set $(\tilde{p},\hat{p})$, the phase angle to maximize the above equation is
\begin{equation}
    \theta_{k}^{\ast}(\tilde{p},\hat{p}) = -\angle{d_{n_{\text{t}},k}^{(\tilde{p})}g_{k,n_{\text{r}}}^{(\hat{p})}}.
\end{equation}
Amongst $L_{d}L_{g}$ candidate paths, we perform a straightforward exhaustive search to determine the set $(\tilde{p}^{\ast},\hat{p}^{\ast})$ as follows
\begin{equation}
    \left( \tilde{p}^{\ast},\hat{p}^{\ast} \right) =\underset{\tilde{p},\hat{p}}{ \text{arg}\hspace{2mm}\max}\left| d_{n_{{\text{t}}},k}^{\tilde{p}}g_{k,n_{\text{r}}}^{\hat{p}} \right|.
\end{equation}
The $k$-th IRS element's phase angle will then be in the opposite direction of the dominant or the strongest cascaded path of the  BS-IRS-UE link.

\subsubsection{Optimization of an Active Precoding Matrix with Given IRS Phase Shifts}
With the provided IRS RC matrix, we optimize the beamforming matrix $\textbf{W}$ at the BS. The sub-problem in this section can be formulated as
\begin{align}
\label{28}
(P1):\hspace{2.5mm}  & \underset{\textbf{W}}{\max}\hspace{1.5mm}  R \left({\bf{W}}\right)=\sum_{n=0}^{N-1}\log_{2}\left( \frac{\left| \textbf{H}_{n}\textbf{W}_{n}\textbf{W}_{n}^{H}\textbf{H}_{n}^{H} +\sigma^{2}\textbf{I}_{MN_{r}} \right|}{\sigma^{2^{MN_{\mathrm r}}}} \right)\nonumber \\
& \text{s.t.} \hspace{2.5mm}\sum_{n=1}^{N}\left\| \textbf{W}_{n} \right\|^{2}\le P_{\max},
\end{align}
 where $\textbf{H}_{n}$ is the cascaded channel matrix of the $n$-th OFDM symbol of IRS-aided MIMO-OTFS without beamforming matrix \textbf{W}. 
In particular, the sum-of-logarithms problem in $(P1)$ is quite challenging. First, we use the Lagrangian dual transform to remove the ratio parts from the logarithm. As given in (\ref{fp_modified}), by adding auxiliary variables $\boldsymbol{\lambda} = [\lambda_{0},\lambda_{1},...,\lambda_{N-1}]^T$, $(P1)$ can be written as 
\begin{align}
\label{30}
(P2): \hspace{2.5mm} &\underset{\boldsymbol{\gamma}(\textbf{W})}{\max}\hspace{2mm} \hat{{R}}(\boldsymbol{\lambda},\boldsymbol{\gamma})  \nonumber  \\
& \text{s.t.} \hspace{3mm}\sum_{n=0}^{N-1}\left\| \textbf{W}_{n} \right\|^{2}\le P_{\max}.
\end{align}
In (\ref{30}), $\boldsymbol{\gamma} = [\gamma_{0},\gamma_{1},...,\gamma_{N-1}]^T$ is  the set of SNR values  as
\begin{equation}\label{gamma}
    {\gamma_{n}}=\left( \frac{\left| \textbf{H}_{n}\textbf{W}_{n}\textbf{W}_{n}^{H}\textbf{H}_{n}^{H} +\sigma^{2}\textbf{I}_{MN_{r}} \right|}{\sigma^{2^{MN_{\mathrm r}}}} \right), \ n=0,...,N-1.
\end{equation}
The objective function in $(P2)$ can be expressed by following (\ref{fp2}). We use $\left| \textbf{H}_{n}\textbf{W}_{n} +\sigma\textbf{I}_{MN_{r}} \right|$ in place of $\textbf{P}_{m}(\textbf{x})$ and $\sigma^{2^{MN_r}}$ in place of $\textbf{Q}_{m}(\text{x})$ in (\ref{fp2}). TABLE~\ref{tab2} outlines the substitution of the relevant terms of Lemma 1 and Lemma 2. 
\begin{table}[!htbp]
\vspace{-4mm}
     \centering
     \caption{Substitution of the terms of Lemma 1 and Lemma~2.}
  \begin{tabular}{ |m{9em}| c| c|  }
\hline
 Transform & $\textbf{P}_{m}(\text{x}) $& $\textbf{Q}_{m}(\text{x})$  \\ \hline  
Lagrange Dual Transform 
(Lemma 1)& $\left| \textbf{H}_{n}\textbf{W}_{n} +\sigma\textbf{I}_{MN_{r}} \right|$  & $\sigma^{2^{MN_{r}}}$  \\ \hline
Quadratic Transform  (Lemma 2)& $(\textbf{H}_{n}^{H}\textbf{W}_{n}+\sigma\textbf{I}_{MN_{r}})$ & $(\textbf{H}_{n}^{H}\textbf{W}_{n}(\textbf{H}_{n}^{H}\textbf{W}_{n})^{H}$\\
& &$+ 2\sigma^{2^{MN_{r}}})$   \\
 \hline
\end{tabular}
  \label{tab2}
  \vspace{-4mm}
\end{table}
The objective function in $(P2)$ is given by
\begin{align}
\label{31}
\hat{R}(\boldsymbol{\lambda},\boldsymbol{\gamma}) =  \left[\sum_{n=0}^{N-1}\ln(1+\lambda_{n})-
 \sum_{n=0}^{N-1}\lambda_{n}+ \sum_{n=0}^{N-1}\frac{(1+\lambda_{n})\gamma_{n}}{1+{\gamma}_{n}} \right].
\end{align}
Note that $(P1)$ and $(P2)$ are equivalent under the same constraints, i.e., they have the same optimal solution. We maximize (\ref{31}) iteratively. Here, the optimal value of $\boldsymbol{\lambda}$ can be obtained by
\begin{equation}
    \frac{\partial \hat{{R}}(\boldsymbol{\lambda},\boldsymbol{\gamma})}{\partial {\lambda}_{n}} = 0, \hspace{2mm}\forall n.
\end{equation}
The above equation can  be used to determine optimal $\boldsymbol{\lambda}$  (denoted by $\boldsymbol{\lambda}^{\ast}$)
\begin{equation}
    \label{311}
    {\lambda}_{n}^{*} = {\gamma}_{n},\hspace{2mm} \forall n.
\end{equation}
Only the final term of (\ref{31}) is involved in the optimization of beamforming matrix for a fixed $\boldsymbol{\lambda}^{\ast}$. Consequently, the optimization problem $(P2)$ can be reformulated as
\begin{align}
\label{32}
(P3):\hspace{3mm}& \underset{\textbf{W}}{\max}\hspace{2mm}\sum_{n=0}^{N-1}\frac{\tilde{{\lambda}}_{n}\left| \textbf{H}_{n}^{H}\textbf{W}_{n}\textbf{W}_{n}^{H}\textbf{H}_{n}+\sigma^{2}\textbf{I}_{MN_{r}} \right|}{2\sigma^{2^{MN_{\mathrm r}}}+\left| \textbf{H}_{n}^{H}\textbf{W}_{n}\textbf{W}_{n}^{H}\textbf{H}_{n} \right|}   \nonumber   \\
&  \text{s.t.} \hspace{3mm}\sum_{n=0}^{N-1}\left\| \textbf{W}_{n} \right\|^{2}\le P_{\max},
\end{align}
 where $\tilde{{\lambda}}_{n} = 1+{\lambda}_{n}^{*}$. In this respect, $(P3)$ boils down to a maximization problem of sum-of-multiple-ratio, which is still challenging because of the fractional term's complex form. Consequently, in an effort to make the active beamforming design more convenient,  $(P3)$ is redefined by making use of the quadratic transform approach. By adding  auxiliary variable matrices $\bar{\boldsymbol{\alpha}}=\left\{ \boldsymbol{\alpha}_{0},...,\boldsymbol{\alpha}_{N-1} \right\}$ with $\boldsymbol{\alpha}_{n}\in \mathcal{C}^{Mn_{\text{r}} \times Mn_{\text{r}}}$ and $n=0,1...,N-1$, 
  $(P3)$ can be reformulated following (\ref{fp4}) as
\begin{align}
\label{33}
(P4):\hspace{3mm} &\underset{\bar{\boldsymbol{\alpha}},\textbf{W}}{\max}\hspace{3mm}\tilde{R}(\boldsymbol{\lambda}^{*},\bar{\boldsymbol{\alpha}},\textbf{W})  \nonumber  \\
&  \text{s.t.} \hspace{3mm}\sum_{n=0}^{N-1}\left\| \textbf{W}_{n} \right\|^{2}\le P_{\max}.
\end{align}
The objective function in $(P4)$ can be expressed by following (\ref{fp4}). We use $(\textbf{H}_{n}^{H}\textbf{W}_{n}+\sigma\textbf{I}_{MN_{r}})$ in place of $\textbf{P}_{m}(\textbf{x})$ and $(\textbf{H}_{n}^{H}\textbf{W}_{n}(\textbf{H}_{n}^{H}\textbf{W}_{n})^{H}+ 2\sigma^{2^{MN_{r}}})$ in place of $\textbf{Q}_{m}(\text{x})$. Thus, the objective function of $(P4)$ is given by
\begin{align}
\label{34}
\tilde{R}(\boldsymbol{\lambda}^{*},\bar{\boldsymbol{\alpha}},\textbf{W}) = \sum_{n=0}^{N-1}2\sqrt{\tilde{{\lambda}}_{n}}\mathcal{R}({\boldsymbol{\alpha}}_{n}^{H}(\textbf{H}_{n}^{H}\textbf{W}_{n}+\sigma\textbf{I}_{MN_{r}})) -  
\sum_{n=0}^{N-1}\left| \boldsymbol{\alpha}_{n} \right|^{2}(\textbf{H}_{n}^{H}\textbf{W}_{n}(\textbf{H}_{n}^{H}\textbf{W}_{n})^{H}+ 2\sigma^{2^{MN_{r}}}).
\end{align}


To maximize $\tilde{R}(\boldsymbol{\lambda}^{*},\bar{\boldsymbol{\alpha}},\textbf{W})$ iteratively over $\boldsymbol{\alpha}$ and $\textbf{W}$, we set $\frac{\partial (\tilde{R}({\boldsymbol{\lambda}}^{*},\bar{\boldsymbol{\alpha}},\boldsymbol{\gamma}))}{\partial \boldsymbol{\alpha}_{n}} = 0,\hspace{2mm}\forall n$.   This step yields  a closed-form updating rule of ${\boldsymbol{\alpha}}_n$ with a fixed $\textbf{W}$ as
\begin{equation}
\label{35}
\boldsymbol{\alpha}_{n}^{*} = \frac{\sqrt{\tilde{{\lambda}}}_{n}(\textbf{H}_{n}^{H}\textbf{W}_{n}+\sigma\textbf{I}_{MN_{r}})}{2\sigma^{2^{MN_{r}}}+\textbf{H}_{n}^{H}\textbf{W}_{n}(\textbf{H}_{n}^{H}\textbf{W}_{n})^{H}},\hspace{3mm}\forall n.
\end{equation}
With fixed $\bar{\boldsymbol{\alpha}}^{*} =\left\{ \boldsymbol{\alpha}_{0}^{*}, \boldsymbol{\alpha}_{1}^{*}, \ldots, \boldsymbol{\alpha}_{N-1}^{*} \right\}$ and neglecting the constant term, we can further write (\ref{34}) as
\begin{equation}
\label{36}
\tilde{R}(\boldsymbol{\lambda}^{*},\bar{\boldsymbol{\alpha}}^{*},\textbf{W}) = -\textbf{W}^{H}\boldsymbol{\mu}\textbf{W} + {\text{Re}}(2\boldsymbol{\rho}^{H}\textbf{W}),
\end{equation}
where
\begin{equation}
\label{37}
\boldsymbol{\mu} = \sum_{n=0}^{N-1}\textbf{H}_{n}\boldsymbol{\alpha}_{n}^{*}\boldsymbol{\alpha}_{n}^{*^{H}}\textbf{H}_{n}^{H} \quad \text{and} \quad \boldsymbol{\rho} = \textbf{H}_{n}\boldsymbol{\alpha}_{n}^{*}.
\end{equation}
The beamforming matrix optimization problem is expressed as the  quadratically constrained convex quadratic program form as shown below
\begin{align}
\label{39}
(P5):\hspace{3mm}& \underset{\textbf{W}}{\min}\hspace{3mm}g(\textbf{W}) = \textbf{W}^{H}\boldsymbol{\mu}\textbf{W} - 2\mathcal{R}\left\{ \boldsymbol{\rho}^{H}\textbf{W} \right\}  \nonumber   \\
& \text{s.t.} \hspace{3mm} \textbf{W}^{H}\textbf{W}\le P_{\mathrm{max}}.
\end{align}
Now the above objective function can be solved using numerous existing techniques. We use a convex optimization tool to obtain the optimal result.

Algorithm 1  summarizes the main steps of the proposed method.
\begin{algorithm}
\caption{Optimization of the beamforming matrix}\label{alg:cap}
\textbf{Input}: Initialize \textbf{W} randomly s.t. $\textbf{W}^{H}\textbf{W}\le P_{\max}$\newline
\textbf{Repeat} \newline
 1: update $\boldsymbol{\lambda}^{*}$ by (\ref{311})\newline
 2: update $\boldsymbol{\bar{\alpha}}^{*}$ by (\ref{35})\newline
 3: update $\textbf{W}$ by (\ref{39}) \newline
 \textbf{Until} problem ($P2$) in (\ref{31}) converges
\end{algorithm}

\vspace{-6mm}
\section{Detection using ADMM}\label{sec4}
The traditional linear detectors are MMSE and zero-forcing (ZF)  and they have decent detection performance. 
 However, a primary challenge with these detectors lies in the matrix inversion operation during the detection process.
 To address this issue, several low-complexity methods have been proposed.
 However, the reduction in computational complexity of these techniques comes at the expense of poorer detection performance.
In contrast, the maximum likelihood (ML) detector, which is a nonlinear detector, is the most prominent one \cite{ml1}.
It can achieve ideal detection performance but suffers from exponentially increasing computational complexity with a growing number of antennas.
The effective channel matrix of the IRS-OTFS system is obtained from the product of the sparse channel matrices of the BS-IRS and IRS-UE links. This product destroys the sparsity and makes the MP detector unsuitable for IRS-OTFS detection. 
Hence, we consider the ADMM detection technique. The  ADMM approach is suitable for convex and non-convex problems owing to its ease of use, operator splitting capabilities, and assured convergence under normal conditions \cite{admm_1}. As shown in the results, the ADMM detector outperforms traditional detectors in terms of bit error rate (BER) performance.

\vspace{-4mm}
\subsection{ADMM Detection}
For signal detection in downlink MIMO systems, where the UE and the BS are equipped with $N_{\text{r}}$ and $N_{\text{t}}$  antennas respectively, the received signal is written as \vspace{-4mm}
\begin{equation}
    \textbf{y}^{\mathrm{DD}} = \textbf{H}_{b}\textbf{a}^{\mathrm{DD}} + \textbf{z},
\end{equation}
where  $\textbf{H}_{b}$ is the channel matrix of IRS-based MIMO OTFS system. Note that $\textbf{H}_{b}$  is the effective channel matrix which includes beamforming matrix as given in (\ref{H4}),    $\textbf{a}^{\mathrm{DD}} \in \mathcal{A}^{N_{\text{t}}MN}$ is modulation symbol.  $\mathcal{A}$ denotes the signal constellation as  $\mathcal{A} = \left\{ a= a_{R} + ja_{I}|a_{R},a_{I}\in \left\{ \pm1 ,\pm3,...,(\pm2^{Q}-1)\right\} \right\}$ and $Q$ is a positive integer. The ideal ML detector for $4^{Q}$-QAM signals can be expressed as the optimization problem described below \cite{admm1}
\begin{equation}\label{ad1}
    \underset{\textbf{a}^{\mathrm{DD}}\in \mathcal{A}^{N_{\text{t}}MN}}{\min}\hspace{2mm}\left\| \textbf{y}^{\mathrm{DD}}-\textbf{H}_{b}\textbf{a}^{\mathrm{DD}} \right\|_{2}^{2}.
\end{equation}
It is practically impossible to obtain its global solution 
due to the exponentially increasing computing complexities associated with the number of antennas on the BS and the size of the set $\mathcal{A}$ \cite{admm2}. 
Higher-order QAM symbols are broken down into a collection of several binary variables, allowing us to express $\textbf{a}^{\mathrm{DD}}$ as \vspace{-4mm}
\begin{equation}
    \textbf{a}^{\mathrm{DD}} = \sum_{q=1}^{Q}2^{q-1}\textbf{a}_{q}^{\mathrm{DD}},
\end{equation}
where $\textbf{a}_{q}^{\mathrm{DD}} \in \mathcal{A}_{q}^{N_{\text{t}}MN}$ and $\mathcal{A}_{q} = \left\{ a_{q}= a_{qR} + ja_{qI}|a_{qR},a_{qI}\in \left\{ +1,-1\right\} \right\}$. 
Now, we can write (\ref{ad1}) as \vspace{-3mm}
\begin{equation}\label{ad2}
    \underset{\textbf{a}_{q}^{\mathrm{DD}}}{\min}\hspace{2mm}\frac{1}{2}\left\| \textbf{y}^{\mathrm{DD}} -\textbf{H}_{b}\sum_{q=1}^{Q}2^{q-1}\textbf{a}_{q}^{\mathrm{DD}}\right\|_{2}^{2}.
\end{equation}
Instead of the binary constrained alphabet  $\mathcal{A}_{q} = \left\{ a_{q}= a_{qR} + ja_{qI}|a_{qR},a_{qI}\in \left\{ +1,-1\right\} \right\}$, we consider box-constrained alphabet $\tilde{\mathcal{A}}_{q} = \left\{ a_{qR} +ja_{qI}|a_{qR},a_{qI}\in\left[ 1,-1 \right]\right\}$.  Adding the sum of the regularisation functions to  (\ref{ad2}), the detection problem can be formulated as
\begin{align}\label{ad3}
   & \underset{\textbf{a}_{q}^{\mathrm{DD}}} {\min}\hspace{2mm}\frac{1}{2}\left\| \textbf{y}^{\mathrm{DD}} - \textbf{H}_{b}\left(\sum_{q=1}^{Q}2^{q-1}\textbf{a}_{q}^{\mathrm{DD}}\right )\right\|_{2}^{2} - \sum_{q=1}^{Q}\frac{\alpha_{q}}{2}\left\| \textbf{a} _{q}^{\mathrm{DD}}\right\|^{2}_{2}  \nonumber \\
    & \text{s.t.} \hspace{3mm} \textbf{a}_{q}^{\mathrm{DD}}\in \tilde{\mathcal{A}}_{q}^{N_{\text{t}}MN}, \hspace{3mm} q=1,..,Q.
\end{align}
 where $\alpha_{q}>0$ is the penalty parameter. 
\vspace{-4mm}
\subsection{\label{admm2} Description of the Detection Algorithm }
In this subsection, we provide a detailed explanation of the ADMM-based detection process for IRS-OTFS. 
By   adding auxiliary variables $\textbf{a}_{0}$,  we  convert (\ref{ad3})  into the following \cite{admm3}
\begin{align}\label{ad4}
    \underset{\textbf{a}_{0},\textbf{a}_{q}^{\mathrm{DD}}}{\min}\hspace{3mm}&\frac{1}{2}\left\| \textbf{y}^{\mathrm{DD}} -\textbf{H}_{b}\textbf{a}_{0}\right\|_{2}^{2}-\sum_{q=1}^{Q}\frac{\alpha_{q}}{2}\left\| \textbf{a}_{q}^{\mathrm{DD}} \right\|_{2}^{2}   \nonumber \\
     \text{s.t.} &\hspace{3mm} \textbf{a}_{0} = \sum_{q=1}^{Q}2^{q-1}\textbf{a}_{q}^{\mathrm{DD}},  \quad \text{and} \quad 
    \textbf{a}_{q}^{\mathrm{DD}}\in \tilde{\mathcal{A}}_{q}^{N_{t}MN},\hspace{3mm} q=1,...,Q.  
   \end{align}
The augmented Lagrangian of  the problem (\ref{ad4}) in simplified form  has the following expression
 \begin{align}\label{ad5}
  & L_{P}(\left\{ \textbf{a}_{q}^{\mathrm{DD}} \right\}_{q=1}^{Q},\textbf{a}_{0},\boldsymbol{\epsilon}) = \frac{1}{2}\left\| \textbf{y}^{\mathrm{DD}}-\textbf{H}_{b}\textbf{a}_{0} \right\|_{2}^{2} -\sum_{q=1}^{Q}\frac{\alpha_{q}}{2}\left\| \textbf{a}_{q}^{\mathrm{DD}} \right\|_{2}^{2} \nonumber \\ &+{\text{Re}}\left\{\boldsymbol{\epsilon}^{T}\left(\textbf{a}_{0}-\sum_{q=1}^{Q}2^{q-1}\textbf{a}_{q}^{\mathrm{DD}}\right )  \right\}  
  +\frac{\rho}{2}\left\| \textbf{a}_{0}-\sum_{q=1}^{Q}2^{q-1}\textbf{a}_{q}^{\mathrm{DD}} \right\|_{2}^{2},
 \end{align}
where $\rho > 0$ and $\boldsymbol{\epsilon}$ are the penalty parameter and Lagrangian multiplier, respectively. 
The framework of the ADMM algorithm using the augmented Lagrangian method is given by
\begin{equation}\label{ad6}
    (\textbf{a}_{q}^{\mathrm{DD}})^{i+1} = \underset{\textbf{a}_{q}^{\mathrm{DD}}\in \tilde{\mathcal{A}}^{N_{t}MN}}{{\text{arg}}\hspace{2mm} \text{min}}\hspace{2mm}L_{p}(\left\{ \textbf{a}_{q}^{\mathrm{DD}} \right\}_{q=1}^{Q},\textbf{a}_{0}^{i},\boldsymbol{\epsilon}^{i}),
\end{equation}
\begin{equation}\label{ad7}
    \textbf{a}_{0}^{i+1} = \underset{\textbf{a}_{0}}{{\text{arg}}\hspace{2mm} {\text{min}}}\hspace{2mm}L_{p}((\textbf{a}_{q}^{\mathrm{DD}})^{i+1},\textbf{a}_{0},\boldsymbol{\epsilon}^{i}), 
\end{equation}
  \begin{equation}\label{ad8}
     \boldsymbol{\epsilon}^{i+1} = \boldsymbol{\epsilon}^{i} + \rho\left(\textbf{a}_{0}^{i+1}-\sum_{q=1}^{Q}2^{q-1}(\textbf{a}_{q}^{\mathrm{DD}})^{i+1}\right), 
  \end{equation}  
where $i$ is the iteration index.\newline
The main challenge lies in finding effective methods to solve the optimization sub-problems (\ref{ad6}) and (\ref{ad7}).  In these sub-problems $L_{p}(\textbf{a}_{q}^{\mathrm{DD}},\textbf{a}_{0}^{i},\boldsymbol{\epsilon}^{i})$  represents a  convex quadratic function with respect to $\textbf{a}_{q}^{\text{DD}}$. Now the gradient of the  augmented Lagrangian function with respect to $\textbf{a}_{q}^{\mathrm{DD}}$  is equated to zero, i.e.,
\begin{equation}\label{ad9}
    \bigtriangledown_{\textbf{a}_{q}^{\mathrm{DD}}}L_{p}\left(\left\{ \textbf{a}_{q}^{\mathrm{DD}} \right\}_{q=1}^{Q},\textbf{a}_{0}^{i},\boldsymbol{\epsilon}^{i}\right) = {\bf{0}}.
\end{equation}
To simplify the analysis, we assume  $ Q= 1$. Solving  (\ref{ad9}) for $Q=1$, we obtain
\begin{equation}\label{57}
    (\textbf{a}_{1}^{\mathrm{DD}})^{i+1} = \underset{[-1,1]}{\Pi}\left( \frac{1}{\rho-\alpha}\left( \rho\textbf{a}_{0}^{i}+{\boldsymbol{\epsilon}}^{i}\right) \right),
\end{equation}
where the operator $\underset{[-1,1]}{\Pi}\left( \cdot  \right)$ projects each entry's real and imaginary components of the given input vector onto the interval $[-1,1]$. We set $\bigtriangledown_{\textbf{a}_{0}} L_{p}\left( (\textbf{a}_{1}^{\mathrm{DD}})^{i+1},\textbf{a}_{0},\boldsymbol{\epsilon}^{i} \right)$ to zero, which leads to the optimal solution for the sub-problem (\ref{ad7}). Solving the associated linear equation, we obtain
 \begin{equation}\label{58}
     \textbf{a}_{0}^{i+1} = (\textbf{H}_{b}^{H}\textbf{H}_{b}+\rho \textbf{I})^{-1}(\textbf{H}_{b}^{H}\textbf{y}+\rho(\textbf{a}_{1}^{\mathrm{DD}})^{i+1}-\boldsymbol{\epsilon}^{i}).
 \end{equation}
Utilizing the penalty-sharing ADMM, we perform detection in the IRS-aided OTFS system. Further, Algorithm \ref{alg:cap} summarizes the main steps of the proposed ADMM detector along with the beamforming.


\vspace{-4mm}
\begin{algorithm}
\caption{ ADMM algorithm for the detection  in the IRS-OTFS system}\label{alg:cap}
\textbf{Inputs}: $ \textbf{H}_{b}$,$N$,$M$, $\rho$, $\alpha$ \newline
\textbf{Output}: $\textbf{a}_{0}^{i}$ \newline
\textbf{Detection}:\newline
\textbf{Initialization}:\newline $(\textbf{a}_{1}^{\mathrm{DD}})^{1}$, $\textbf{a}_{0}^{1}$, $\boldsymbol{\epsilon} $ = $\boldsymbol{0}$ vector\newline
\textbf{Detection}:\newline
$i$ = 1  \newline
\textbf{while}  \hspace{0.5mm}  Stopping criteria are not met \newline
1: update $(\textbf{a}_{1}^{\mathrm{DD}})^{i+1}$ using (\ref{57})\newline
2: Update $\textbf{a}_{0}^{i+1}$ using (\ref{58})\newline
3: Update $\boldsymbol{\epsilon}^{i+1}$ using (\ref{ad8})\newline
$i=i+1$ \newline
\textbf{do}
\end{algorithm}
\vspace{-4mm}

 \subsection{Complexity Analysis}
Computational complexity of the signal detection using the ADMM-based approach is summarized. The practical feasibility of the detection algorithm relies on its computational complexity, particularly the number of complex multiplications involved.
The total computational complexity of the ADMM detector algorithm is divided into two parts: the iteration-independent part and the iteration-dependent part. The iteration-independent part is executed only once. Iteration-independent  computation involves three steps: ${\textbf{H}_{b}^H}\textbf{H}_{b}$, $(\textbf{H}_{b}^{H}\textbf{H}_{b}+\rho \textbf{I})^{-1}$  and $\textbf{H}_{b}^{H}\textbf{y}$. In these three steps, the number of required complex multiplications are $(N_{\mathrm r}NM)(N_{\mathrm t}NM)^{2}$, $(N_{\mathrm t}NM)^{3}$ and $(N_{\mathrm r}NM)(N_{\mathrm t}NM)$.
The iteration-dependent section must be repeated for each iteration and primarily consists of two steps. These steps involve scalar multiplication of a $(N_{\mathrm t}NM \times 1)$   vector for $\textbf{a}_{1}$ update in (\ref{57}) and multiplication of a $(N_{\mathrm t}NM \times N_{\mathrm t}NM)$  matrix with a $(N_{\mathrm t}NM \times 1) $ vector and scalar multiplication with another $(N_{\mathrm t}NM \times 1)$  vector in (\ref{58}).
 \begin{table}[!htbp]
     \centering
     \caption{Complexity Analysis.}
\begin{tabular}{ |c| c| }
\hline
 Algorithm & Computational complexity \\ \hline  \hline
ADMM  & $(N_{t}NM)^{3}+(N_{r}NM)(N_{t}NM)^{2}+(N_{r}NM)(N_{t}NM)$ \\
& $+i(N_{t}NM+(N_{t}NM)^{2}+N_{t}NM)$\\ \hline
MMSE & $(N_{t}NM)^{3}+(N_{r}NM)(N_{t}NM)^{2}+(N_{r}NM)(N_{t}NM)$\\
\hline
\end{tabular}
  \label{tab3}
\end{table}
Additionally, we include the overall complexity of the traditional linear MMSE \cite{admm3}. For MMSE, only these three steps are required for computation: $\textbf{H}_{b}^{H}\textbf{H}_{b}$, $(\textbf{H}_{b}^{H}\textbf{H}_{b}+\rho \textbf{I})^{-1}$  and $\textbf{H}_{b}^{H}\textbf{y}$, and the number of required complex multiplications are $(N_{\mathrm r}NM)(N_{\mathrm t}NM)^{2}$, $(N_{\mathrm t}NM)^{3}$ and $(N_{\mathrm r}NM)(N_{\mathrm t}NM)$.
 In conclusion, the ADMM algorithm has a  higher computational complexity than linear algorithms but lower computational complexity than non-linear methods but it outperforms the linear ones as shown in the results.
\section {Channel Estimation of IRS-OTFS using Tensor Modelling}\label{chaest}
The reliability of the channel state information determines the potential benefits of the IRS-OTFS system. In order to solve the channel estimation problem, a tensor modelling approach is used for tackling the receiver design of the IRS-OTFS system in this section. A channel estimate technique based on a parallel factor (PARAFAC) tensor analysis for the received signals is provided, taking into account an organised time-domain sequence of pilots and IRS phase changes. 
\subsection{ ALS Channel Estimation}
Following the effective channel explained in Section II and considering only cascaded path, the cascaded channel matrix of the OTFS frame can be written as: $\textbf{H}_{\text{cas}} = \textbf{D}\Theta\textbf{G}$. The IRS is assumed to modify its phase-shifts in a time-slotted transmission based on the time $t=1, \cdots,T$. The IRS-User and BS-IRS channels are assumed to be continuous over $T$ time slots. The received signal can be written as
\begin{equation}\label{est1}
\boldsymbol{r}(t) = \textbf{D}(\text{diag}(\boldsymbol{\theta}(t))\otimes \textbf{I}_{NM})\textbf{G}\boldsymbol{x}(t) +\textbf{z}(t),
\end{equation}
where the vector $x(t)\in \mathcal{C}^{N_{t}NM \times 1} $ at time $t$ contains the pilot signals that were transmitted.
In (\ref{est1}),   $\boldsymbol{\theta}(t) = (\beta_{1,t}e^{j\theta_{1}},\cdots ,\beta_{k,t}e^{j\theta_{k}})$.
 $\boldsymbol{\theta}(t) $ refers to  the IRS's phase shift at time $t$. The channel coherence time $T_{s}$ is divided into $L$ blocks, with $T$ time slots in each block, resulting in $T_{s} = LT$ \cite{ch1}. The signal received in the $t$-th time frame of the $l$-th block is $\textbf{r}[l,t] = r[(l-1)T+t]$, $t = 1,2, \cdots,T, l = 1,2, \cdots,L$. The phase shift vectors  and pilot signal connected to the $t$-th time frame of the $l$-th block are represented by the notations  $\boldsymbol{\theta}[l,t]$ and $\textbf{x}[l,t]$, respectively. For all $L$ blocks, $\boldsymbol{\theta}[l,t]$ varies, but for $T$ time slots, it is constant. Over $L$ blocks, $\textbf{x}[l,t]$ is repeated \cite{ch1}. Mathematically,
\begin{equation}\label{est2}
  \boldsymbol{\theta}[l,t] =  \boldsymbol{\theta}[l],\hspace{2mm} \text{for} \hspace{2mm} t = 1, \cdots,T, \ \text{and} \  \textbf{x}[l,t] = \textbf{x}[t],\hspace{2mm} \text{for} \hspace{2mm} l = 1,\cdots,L.
\end{equation}
These presumptions allow the received signal (\ref{est1}) to be expressed as:
\begin{equation}
    \textbf{r}[l,t] = \textbf{D}(\text{diag}(\boldsymbol{\theta}[l])\otimes \textbf{I}_{NM})\textbf{G}\textbf{x}[t]+\textbf{z}[l,t].
\end{equation}
The $l$-th block in \textbf{R}$[l]$ gathers all the signals received within the $T$ time frame
\begin{equation}\label{est4}
 \textbf{R}[l] = [\textbf{r}[l,1], \cdots,\textbf{r}[l,T]]\in \mathcal{C}^{N_{\text{r}}NM \times T}.   
\end{equation}
In matrix form, we get
\begin{equation}\label{est5}
\textbf{R}[l] = \textbf{D}(\text{diag}(\boldsymbol{\theta}[l])\otimes \textbf{I}_{NM})\textbf{G}\textbf{X}^{T}+\textbf{Z}[l]
.\end{equation}
For simplification, the noise term is removed and (\ref{est5}) can be written as
\begin{equation}\label{est6}
    \bar{\textbf{R}}[l] = \textbf{D}(\text{diag}(\boldsymbol{\theta}[l])\otimes \textbf{I}_{NM})\textbf{B}^{T}, \hspace{3mm}\textbf{B} = \textbf{X}\textbf{G}^{T}.
\end{equation}
One method to define the matrix $\bar{\textbf{R}}[l]$ is as the $l$-th frontal matrix layer of a three-way tensor $\bar{\textbf{r}} \in \mathcal{C}^{N_{\text{r}}NM\times T \times L}$ that undergoes a PARAFAC decomposition and considering $\textbf{D}_{l}(\boldsymbol{\phi})  = \text{diag}(\boldsymbol{\theta}(l))$. And $\boldsymbol{\phi} = [\boldsymbol{\theta}[1],....,\boldsymbol{\theta}[L]]^T$. $\textbf{D}_{l}(\boldsymbol{\phi}) $ represents a diagonal matrix with its main diagonal including the $l$-th row of $\boldsymbol{\phi}$. Following PARAFAC decomposition, the received signal tensor $\bar{\textbf{r}}$ can be split into these three matrices \cite{ch1} as
\begin{equation}\label{est7}
\textbf{R}_{1} = \textbf{D}[(\boldsymbol{\phi}\otimes \textbf{I}_{NM}) \bullet \textbf{B} ]^{T},  \quad \textbf{R}_{2} = \textbf{B}[(\boldsymbol{\phi}\otimes \textbf{I}_{NM}) \bullet \textbf{D} ]^{T}, \quad \text{and} \  \textbf{R}_{3} = (\boldsymbol{\phi}\otimes \textbf{I}_{NM})[\textbf{B} \bullet \textbf{D} ]^{T},
\end{equation}
where $\bullet$ represents Khatri Rao product, i.e., matching column wise kronecker product.
From (\ref{est7}), an alternating least squares approach can be used to obtain an iterative solution. As $\boldsymbol{\phi}$ is available at the receiver in our scenario and we have  initialized \textbf{D} and \textbf{G} with all zeros, the process includes iteratively optimizing the subsequent two cost functions to estimate the matrices \textbf{D} and \textbf{G} as shown below
\begin{equation}\label{estnew}
    \hat{\textbf{D}} = \underset{\textbf{D}}{\arg\hspace{2mm}\min}\left\| \textbf{R}_{1}-\textbf{D}((\boldsymbol{\phi}\otimes \boldsymbol{I}_{NM})\bullet \textbf{X}\textbf{G}^{T})^{T} \right\|_{F}
\end{equation}
\begin{equation}\label{est9}
    \hat{\textbf{G}} = \underset{\textbf{G}}{\arg\hspace{2mm}\min}\left\| \textbf{R}_{2}-\textbf{X}\textbf{G}^{T}((\boldsymbol{\phi}\otimes \boldsymbol{I}_{NM})\bullet \textbf{D})^{T} \right\|^{2}_{F}
\end{equation}
which are, in turn, solved by \vspace{-2mm}
\begin{equation}\label{est10}
    \hat{\textbf{D}} = \textbf{R}_{1}[((\boldsymbol{\phi}\otimes \textbf{I}_{NM})\bullet \textbf{X}\textbf{G}^{T})^{\dagger} ] \ \text{and} \  \hat{\textbf{G}} = \textbf{X}^{\dagger}\textbf{R}_{2}[((\boldsymbol{\phi}\otimes \textbf{I}_{NM})\bullet \textbf{D})^{T}]^{\dagger}.
\end{equation}
where $\textbf{X}^{\dagger}$ represents the pseudo-inverse of the  matrix \textbf{X}.

 \vspace{-4mm}
 \section{Simulation Results}\label{sec5}
 This section presents the sum rate and BER analysis of the proposed IRS-aided OTFS system with beamforming and low complexity ADMM-based detector\footnote{ Sample simulation codes for this work are available in  the following link: https://github.com/Sushmita898/IRS-Assisted-OTFS-Beamforming-Design-and-Signal-Detection-Joint-Beamforming    https://github.com/Sushmita898/IRS-Assisted-OTFS-Beamforming-Design-and-Signal-Detection}. The transmitted symbols in the DD grid are QAM modulated. The simulations are carried out for two propagation models:
 
     (1) {\textit{Basic Propagation Model (BPM):-}} 
     The simple propagation environments of BPM are taken as the first  to understand the behavior of the IRS-OTFS system. We take into account an IRS-aided OTFS network with $L_{g} = L_{h} = L_{d} = 4$ and a DD grid size of $N=32$, $M=32$  for simulations.  TABLE~\ref{tabnew1} summarizes the DD-domain parameters and other simulation details for BPM.
\begin{table}[!htb]
    \caption{BPM and EVA Channel Models' Simulation Parameters.}
    \begin{center}
        \subtable{
            \begin{tabular}{|c|p{1.5in}|}
                \hline
                \multicolumn{2}{|c|}{Simulation Parameters for BPM} \\  \hline \hline
                \multicolumn{1}{|c|}{Item} & \multicolumn{1}{c|}{Values} \\  \hline
Subcarrier spacing, $\Delta f$  & 15 KHz\\
Carrier frequency, $f_{c}$ & 4 GHz  \\
Total Doppler bins, $N$ & 32 \\
Total delay bins, $M$ & 32 \\
Number of IRS element, $K$ & 16\\ 
Number of paths for direct link, $L_{h}$& 4\\
Number of paths for BS-IRS link, $L_{d}$ & 4\\
Number of paths for IRS-UE,$L_{g}$ & 4\\
Delay tap & Randomly from $\left\{0,1,..,3\right\}$\\
Doppler tap & Randomly from $\left\{0,1,..,3\right\}$\\
                \hline
            \end{tabular}
            \centering
        } 
        \subtable{
          \centering
            \begin{tabular}{|p{1.5in}|p{.62in}|}
                \hline
                \multicolumn{2}{|c|}{Simulation Parameters for EVA Model} \\
                \hline \hline
                \multicolumn{1}{|c|}{Item} & \multicolumn{1}{c|}{Values} \\
                \hline  \hline
Subcarrier spacing, $\Delta f$  & 15 KHz\\
Carrier frequency, $f_{c}$ & 4 GHz  \\
Number of Doppler bins, $N$ & 16 \\
Number  of  delay bins, $M$ & 512 \\
Number of paths for direct link, $L_{h}$& 9\\
Number of paths for BS-IRS link, $L_{d}$ & 9\\
Number of paths for IRS-UE,$L_{g}$ & 9\\
                \hline
            \end{tabular}   
        } 
    \end{center}
    \label{tabnew1}
\end{table}

(2) {\textit{Extended Vehicular A (EVA) Propagation Model:-}} The EVA propagation model is more comprehensive and practical for a high-mobility scenario \cite{eva_par}. To analyze the performance of the IRS-OTFS system, the EVA propagation environment is considered. We consider a scenario  with $L_{g} = L_{h} = L_{d} = 9$ and  OTFS grid dimension of $N=16$ and $M=512$.  The subcarrier spacing is taken as 15 KHz. The maximum integer delay spread is considered to be 20. The typical EVA model is used to generate the delay taps of the channel. The excess path delays are  [0, 30, 150, 310, 370, 710, 1090, 1730, 2510] ns with a power delay profile of [0, -1.5, -1.4, -3.6, -0.6, -9.1, -7.0, -12, -16.9] dB \cite{eva_par}. A uniform distribution $\mathcal{U}(0, \nu_{\text{max}})$ is used to generate the Doppler shifts where $\nu_{\text{max}}$  is the most significant Doppler shift determined by the speed of the vehicle. TABLE~\ref{tabnew1} shows the simulation parameters for the EVA propagation model.


The amplitudes of the channel coefficients are assumed to follow Rayleigh distribution. For the multiple-input single-output (MISO) and MIMO configurations, we take $N_{\text{t}}=2, N_{\text{r}}=1 $  and  $N_{\text{t}}=2, N_{\text{r}}=2 $, respectively.    The sum rate and the BER performance of the IRS-aided OTFS system for both BPM and EVA propagation models are analyzed in the subsequent subsections.
\vspace{-4mm}
 \subsection{Sum Rate Analysis}
\begin {figure}[!htbp]
\vspace{-4mm}
\centering
	\includegraphics[scale =0.5]{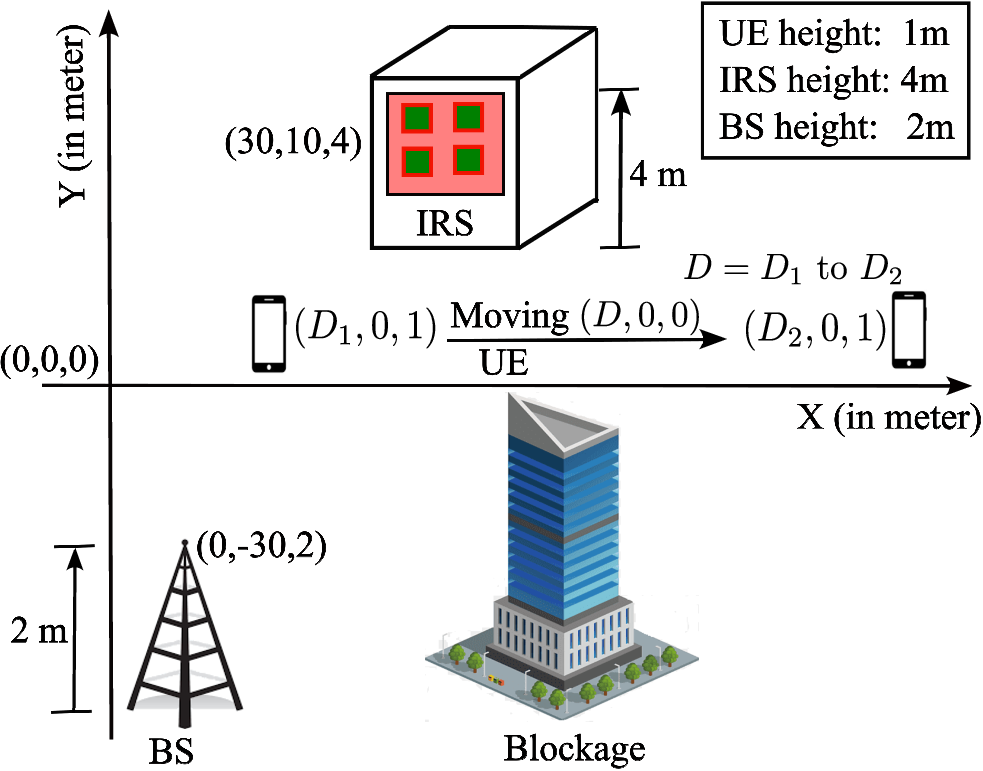}
 \vspace{-2mm}
	\caption{\footnotesize{ The simulation environment where BS assisted by IRS serve single moving UE. }}
	\label{figsim}
 \vspace{-4mm}
\end{figure}

This subsection of the paper demonstrates the theoretical foundation of the sum rate of the IRS-OTFS system.
 A 3D scenario is considered for simulation, as shown in Fig~\ref{figsim}. In this setup, a single BS serves a single user, and the network's capacity is limited by scatterers' interference. A single IRS  is deployed on a tall building surface to increase the capacity. 
The location of the BS is considered as (0 m, -30 m, 2 m), while the IRS is located at (30 m, 10 m, 4~ m). Initially, the MISO  case is considered for simplicity, and then we extend it to the MIMO case.
 The SNR is specified as  $\frac{\textbf{E}\left( \left| a[l,k] \right| ^{2}\right)}{\sigma_{n}^{2}}$, where $\sigma_{n}^{2}$ represents the noise variance. Both the cascaded link and the direct link are taken into account unless otherwise specified.

\pgfplotsset{every semilogy axis/.append style={
		line width=0.9 pt, tick style={line width=0.9pt}}, width=8cm,height=6.5cm, 
	legend style={font=\scriptsize},
	legend pos= north west}
\begin{figure}
\centering
\begin{minipage}{.5\textwidth}
  \centering
  \centering
		\begin{tikzpicture}[new spy style]
			\begin{axis}[xmin=-15, xmax=5, ymin=0, ymax=7,
				xlabel={\small SNR (dB)},
				ylabel={\small Sum rate (bit/s/Hz)},
				grid=both,
				grid style={dotted},
				legend cell align=left,
				legend entries={Case 3,Case 1,Case 4,Case 2,Case 5, Case 6},
				cycle list name=mycolorlist,
				]
				\node[text width=3cm] at (axis cs:2,6.4){\scriptsize $N=32$ };
    \node[text width=3cm] at (axis cs:2,5.9){\scriptsize $M=32$ };
    \node[text width=3cm] at (axis cs:2,5.4){\scriptsize $K=16$ };
	\addplot[magenta,dashed,mark=oplus,thick,mark size=3pt, mark options={scale=1,solid}] table [x={x1}, y={y1}] {MISO_cap_comp.txt};
	\addplot [green,mark=o,thick,mark size=3pt] table [x={x1}, y={y2}] {MISO_cap_comp.txt};
    \addplot [blue,mark=pentagon,thick,mark size=3pt]  table [x={x1}, y={y3}] {MISO_cap_comp.txt};
    \addplot [cyan,mark=square,thick,mark size=3pt]  table [x={x1}, y={y4}] {MISO_cap_comp.txt};
    \addplot [red,mark=triangle,thick,mark size=3pt] table [x={x1}, y={y5}] {MISO_cap_comp.txt};           
     \addplot [black,mark=star,thick,mark size=3pt] table [x={x1}, y={y6}] {MISO_cap_comp.txt};           
			\end{axis}
		\end{tikzpicture}		
		\vspace*{-0.1cm}
		\caption{\footnotesize{ Sum rate against SNR for MISO IRS OTFS system in BPM.}}	
	\label{miso}
\end{minipage}%
\begin{minipage}{.5\textwidth}
  \centering
  \begin{tikzpicture}[new spy style]
			\begin{axis}[xmin=-15, xmax=5, ymin=0, ymax=7,
				xlabel={\small SNR (dB)},
				ylabel={\small sum rate(bit/s/Hz)},
				grid=both,
				grid style={dotted},
				legend cell align=left,
         legend entries={ Case 3, Case 1,Case 4,Case 2, Case 5, Case 6, Case 7, IRS-OFDM},
				cycle list name=mycolorlist,
				]
				\node[text width=3cm] at (axis cs:-1,4.8){\scriptsize $K=16$ };
    \node[text width=3cm] at (axis cs:-1,5.8){\scriptsize $M=32$ };
    \node[text width=3cm] at (axis cs:-1,5.3){\scriptsize $N=32$ };
	\addplot [magenta,dashed,mark=oplus,thick,mark size=3pt, mark options={scale=1,solid}] table [x={x1}, y={y1}] {MIMO_cap_comp.txt};
	\addplot [green,mark=o,thick,mark size=3pt]  table [x={x1}, y={y2}] {MIMO_cap_comp.txt};
    \addplot [blue,mark=pentagon,thick,mark size=3pt]  table [x={x1}, y={y3}] {MIMO_cap_comp.txt};
\addplot [cyan,mark=square,thick,mark size=3pt]  table [x={x1}, y={y4}] {MIMO_cap_comp.txt};
\addplot [red,mark=triangle,thick,mark size=3pt]  table [x={x1}, y={y5}] {MIMO_cap_comp.txt};
    \addplot [black,mark=star,thick,mark size=3pt]  table [x={x1}, y={y6}] {MIMO_cap_comp.txt}; 
     \addplot [orange,mark=+,thick,mark size=3pt]  table [x={x1}, y={y7}] {MIMO_cap_comp.txt};
      \addplot [blue,dashed,mark=triangle,thick,mark size=3pt]  table [x={x1}, y={y8}] {MIMO_cap_comp.txt};
			\end{axis}
		\end{tikzpicture}		
		\vspace*{-0.1cm}
		\caption{\footnotesize{ Sum rate against SNR for MIMO IRS-OTFS system in BPM.}	}	
	\label{mimo}
\end{minipage}
\vspace{-4mm}
\end{figure}

\subsubsection{Achievable Sum Rate of IRS-OTFS  System for BPM}
In this subsection, we analyze the sum rate of an IRS-OTFS system for BPM. 
We examine the ideal IRS scenario and assume that the UE has a height of $1$~ m.  The number of IRS elements is $K=16$.
Following the simulation parameters for BPM given in TABLE~\ref{tabnew1}, we plot the sum rate against the SNR in Fig. \ref{miso} and Fig.~\ref{mimo} for  MISO and MIMO cases, respectively. 
Using the fully known channel state information (CSI) otherwise specified, we simulate the ASR using the joint precoding framework. The beamforming matrix \textbf{W} is optimized at the BS according to Algorithm~1.  The simulation results are illustrated  through several cases: 
\begin{enumerate}
    \item $Case\hspace{1mm}1:$ The RC phase shift optimization at the IRS is carried out according to the  STM method as described in Section~\ref{sec3}.  
    \item $Case\hspace{1mm}2:$ Here, the  RC phase shifts considered at the IRS are produced from random angles.  
    \item $Case\hspace{1mm}3:$ To compare the obtained results with the ideal scenario, we simulate the achievable sum rate by enforcing coherent phase shifts at the IRS.
    \item  $Case\hspace{1mm}4:$ The RC phase shift optimization at the IRS is carried out according to the  STM  without considering the presence of a D-link .
    \item $Case\hspace{1mm}5:$ Now, we conduct simulations to determine the sum rate for the scenario without an IRS.
     \item $Case\hspace{1mm}6:$ In this case, simulations are carried out for estimated cascaded CSI 
     (imperfect CSI) using tensor modelling approach as discussed in Section V. Here, only cascaded links are taken for simulation and direct link is not considered.
      \item $Case\hspace{1mm}7:$ Simulations are carried out considering Case~1 with fractional Doppler.

\vspace{-4mm}
\begin{table}[!htbp]
     \centering
     \caption{Various Cases for Sum Rate Analysis.}
\begin{tabular}{ |c| c|c|c|c|c| }
\hline
Cases  & IRS RC& D-link&IRS&CSI&Doppler  \\ \hline 
Case 1  & STM & Present & Present & Perfect& Integer \\ \hline
Case 2 & Random &Present&Present &Perfect &Integer \\ \hline
Case 3 & Coherent &Present &Present & Perfect &Integer  \\ \hline
Case 4 & STM & Absent & Present & Perfect &integer \\ \hline
Case 5 & Not applicable& Present & Absent& Perfect &Integer \\

\hline
Case 6 & STM & Absent &Present& Imperfect & Integer \\ \hline
Case 7 & STM & Present & Present & Perfect & Fractional \\ \hline
\end{tabular}
  \label{tab4}
\end{table}
    
  
\end{enumerate}
TABLE~\ref{tab4} outlines the key features of various simulation cases. For all cases, we consider the  total number of IRS elements as $K$ = 16.
 In Fig. \ref{miso} and Fig. \ref{mimo}, we present the sum rates for the IRS-OTFS system against  SNR for both MISO  and MIMO  configurations in the case of BPM. It is noticed that the performance gain is significant for IRS-aided OTFS, particularly when employing the aforementioned proposed algorithm (Case 1). There is a slight decrease in the gain when D-link is not taken into consideration (Case 4). However, the gain is not as prominent as when using random angles (Case~2) due to the fact there is no optimization done at the IRS in this case. Also, the performance of the proposed algorithm (Case 1) approaches the gain of the ideal scenario i.e., coherent phase (Case 3). Further, the performance gain is too low in the absence of IRS (Case 5). Later, the  sum rate for estimated cascaded CSI without D-link is simulated in Case ~6. We address estimated cascaded CSI as imperfect CSI. It is found that the performance with imperfect CSI is poor as compared with perfect CSI. In all of the above cases integer Doppler is considered. Now in Case ~7, fractional Doppler is considered i.e, Case ~1 is incorporated with fractional Doppler instead of integer Doppler. The performance gain in the case of integer Doppler is significant compared  to fractional Doppler  case. Finally, the sum rate for IRS-OFDM is simulated and it is  found that IRS-OTFS outperforms IRS-OFDM.  

\subsubsection{Achievable Sum Rate of IRS-OTFS  System for EVA Propagation Model}
Here, we analyze the achievable sum rate against SNR under the EVA propagation model. The simulation parameters related to the EVA propagation model are shown in TABLE \ref{tabnew1}. For all cases, the total number of IRS elements is  $K$ = 16.
Fig. \ref{mimoeva} shows the achievable sum rate against SNR under the EVA model with various cases presented in TABLE \ref{tab4}. For the EVA model as well, similar observations are obtained. Case~1, i.e., the proposed algorithm, outperforms Cases~2, 4, and 5 in terms of performance gain for the same reasons specified in the BPM scenario. Case 3 has more performance gain than Case 1, yet Case~1  approaches the ideal scenario (Case~3). Similar to BPM, in Case ~6, imperfect cascaded CSI is considered and it is found from simulation that performance gain degrades with imperfect CSI as compared to that with perfect CSI. Later, fractional Doppler is incorporated in Case~1 instead of integer Doppler (Case~7). It is found that the performance gain with integer Doppler is better as compared to the one with fractional Doppler. At last, the sum rate for IRS-OFDM  is far greater than that of IRS-OFDM.  \newline
Furthermore, it is worth discussing that when the transmit power at the BS is too low, the performance gain becomes insignificant. This observation aligns with intuition since, with minimal BS transmit power, the signals reflected by the IRS become weak, resulting in minimal impact from the IRS on performance enhancement.

\pgfplotsset{every semilogy axis/.append style={
		line width=0.9 pt, tick style={line width=0.9pt}}, width=8cm,height=6.5cm, 
	legend style={font=\scriptsize},
	legend pos= north west}
\begin{figure}[htb]
\centering
\begin{minipage}{.47\textwidth}
  \centering
  	\begin{tikzpicture}[new spy style]
			\begin{axis}[xmin=-10, xmax=10, ymin=0, ymax=13,
				xlabel={\small SNR (dB)},
				ylabel={\small Sum rate (bit/s/Hz)},
				grid=both,
				grid style={dotted},
				legend cell align=left,
              legend entries={ Case 3,Case 1,Case 4,Case 2, Case 5, Case 6, Case 7, OFDM IRS},
				cycle list name=mycolorlist,
				]
				\node[text width=3cm] at (axis cs:5,12.0){\scriptsize $N=16$ };
    \node[text width=3cm] at (axis cs:5,11){\scriptsize $M=512$ };
    \node[text width=3cm] at (axis cs:5,10){\scriptsize $K=16$ };
\addplot [magenta,dashed,mark=oplus,thick,mark size=3pt, mark options={scale=1,solid}] table [x={x1}, y={y1}] {MIMO_cap_comp_eva.txt};
\addplot [green,mark=o,thick,mark size=3pt]  table [x={x1}, y={y2}] {MIMO_cap_comp_eva.txt};
\addplot [blue,mark=pentagon,thick,mark size=3pt]  table [x={x1}, y={y3}] {MIMO_cap_comp_eva.txt};
\addplot [cyan,mark=square,thick,mark size=3pt]  table [x={x1}, y={y4}] {MIMO_cap_comp_eva.txt};
\addplot [red,mark=triangle,thick,mark size=3pt]  table [x={x1}, y={y5}] {MIMO_cap_comp_eva.txt};
\addplot [black,mark=star,thick,mark size=3pt]  table [x={x1}, y={y6}] {MIMO_cap_comp_eva.txt};
\addplot [orange,mark=+,thick,mark size=3pt]  table [x={x1}, y={y7}] {MIMO_cap_comp_eva.txt};
\addplot [blue,dashed,mark=hexagon,thick,mark size=3pt]  table [x={x1}, y={y8}] {MIMO_cap_comp_eva.txt};
        \end{axis}
		\end{tikzpicture}		
		\vspace{-8mm}
		\caption{\footnotesize{ Sum rate  for MIMO IRS-OTFS system in EVA Model. }}
	\label{mimoeva}
\end{minipage}%
\begin{minipage}{.48\textwidth}
  \centering
  	\begin{tikzpicture}[new spy style]
			\begin{axis}[xmin=-10, xmax=10, ymin=-.1, ymax=22,
				xlabel={\small SNR (dB)},
				ylabel={\small Sum rate (bit/s/Hz)},
				grid=both,
				grid style={dotted},
				legend cell align=left,
				legend entries={$K$=4 Cae 2,$K$=4 Case 1, $K$=8 Case 2, $K$=8 Case 1, $K$=16 Case 2, $K$=16 Case 1, $K$=32 Case 2,$K$=32 Case 1  },
				cycle list name=mycolorlist,
				]
				\node[text width=3cm] at (axis cs:6,19.5){\scriptsize $N=16$ };
    \node[text width=3cm] at (axis cs:6,18.0){\scriptsize $M=512$ };
   
\addplot [black,dashed,mark=square,thick,mark size=3pt, mark options={scale=1,solid}] table [x={x1}, y={y1}] {MIMO_EVA_wrt_irs_element.txt};
\addplot [magenta,mark=o,thick,mark size=3pt]  table [x={x1}, y={y2}] {MIMO_EVA_wrt_irs_element.txt};
\addplot [blue,mark=pentagon,thick,mark size=3pt]  table [x={x1}, y={y3}] {MIMO_EVA_wrt_irs_element.txt};
\addplot [cyan,mark=diamond*,thick,mark size=3pt] table [x={x1}, y={y4}] {MIMO_EVA_wrt_irs_element.txt};
\addplot [green,mark=oplus,thick,mark size=3pt] table [x={x1}, y={y5}] {MIMO_EVA_wrt_irs_element.txt};
\addplot [blue,dashed,mark=diamond,thick,mark size=3pt, mark options={scale=1,solid}] table [x={x1}, y={y6}] {MIMO_EVA_wrt_irs_element.txt};
\addplot [red,dashed,mark=star,thick,mark size=3pt, mark options={scale=1,solid}] table [x={x1}, y={y7}] {MIMO_EVA_wrt_irs_element.txt};
\addplot [magenta,dashed,mark=oplus,thick,mark size=3pt, mark options={scale=1,solid}] table [x={x1}, y={y8}] {MIMO_EVA_wrt_irs_element.txt};
        \end{axis}
		\end{tikzpicture}		
		\vspace{-8mm}
		\caption{\footnotesize{ Sum rate against SNR (dB) with varying number of IRS elements in EVA Model.}}		
	\label{wrtirs_eva}
\end{minipage}
\vspace{-4mm}
\end{figure}

\subsubsection{Comparison of Sum Rate with Varying Number of IRS Elements}
 Fig. \ref{wrtirs_eva}  presents the sum rate of the MIMO IRS-OTFS system for EVA  with increasing number $K$ of IRS elements.
The performance gain is analyzed for two scenarios: one when the RC matrix is optimized following STM (Case 1) and another when the RC matrix is considered to be random (Case 2). Beamforming matrix optimization in both cases is done following Algorithm~1. It is evident from the results that as the
number of IRS elements increases, the sum rate of the IRS-aided
OTFS system also increases.

\subsubsection{Comparison of Sum Rate w.r.t. Distance}
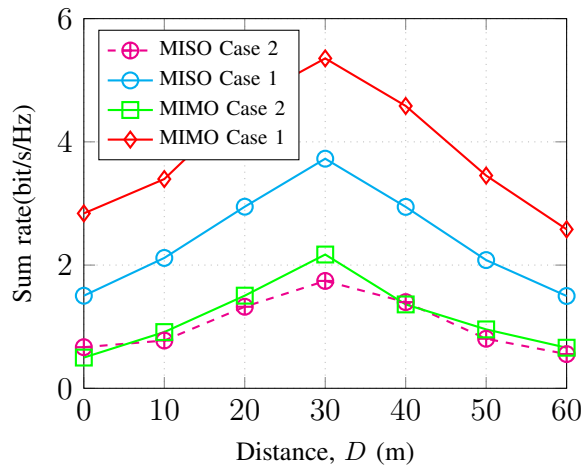
\begin{figure}[!htpb]
		\centering
		\begin{tikzpicture}[new spy style]
			\begin{axis}[xmin=0, xmax=60, ymin=0, ymax=6,
				xlabel={\small Distance, $D$ (m)},
				ylabel={\small Sum rate(bit/s/Hz)},
				grid=both,
				grid style={dotted},
				legend cell align=left,
    legend entries={MISO Case 2, MISO Case 1, MIMO Case 2, MIMO Case 1},
				cycle list name=mycolorlist,
				]
				
\addplot [magenta,dashed,mark=oplus,thick,mark size=3pt, mark options={scale=1,solid}] table [x={x1}, y={y1}] {singleirs_distance.txt};
\addplot [cyan,mark=o,thick,mark size=3pt]  table [x={x1}, y={y2}] {singleirs_distance.txt};
\addplot [green,mark=square,thick,mark size=3pt]  table [x={x1}, y={y3}] {singleirs_distance.txt};
\addplot [red,mark=diamond,thick,mark size=3pt]  table [x={x1}, y={y4}] {singleirs_distance.txt};
               
			\end{axis}
		\end{tikzpicture}		
		\vspace{-4mm}
		\caption{ \footnotesize{ Sum rate in EVA model.}}		
	\label{dist_eva}
 \vspace{-8mm}
\end{figure}

In  Fig.~\ref{dist_eva}, the impact of the user's distance from the IRS  on the system's sum rate for EVA propagation model is analyzed.
Provided that the deployment of the BS, IRS, and user is as shown in Fig. 2, we evaluate the system's performance with regard to the user's position. The total horizontal distance of the BS-IRS-UE link is fixed at $D$ = 60 m and the BS-UE distance is varied. The sum rate is analyzed for various distances from 0 m to 60 m. The results are presented for MISO and MIMO cases. From  Fig.~\ref{dist_eva}, we can clearly observe a peak at a distance of around 30 m in both MISO as well as MIMO cases. Since the user can obtain powerful signals reflected from the IRS, it suggests that the sum rate increases when the users approach the IRS. 
The sum rate for STM-optimized RC phase shift (Case 1) is high for both the MISO and MIMO scenarios. Furthermore, we observe that the ``Random phase angle" (Case 2) yields a very low gain. The reason is that with no RC phase shift optimization, the signals arriving at IRS cannot be precisely guided to the user.

\vspace{-6mm}

\subsection{BER Performance Analysis}
The  BER performances for ADMM and MMSE detectors are compared in  BPM and EVA  propagation models. Various cases are considered, as detailed in TABLE~\ref{tab6}. 



\subsubsection{BER Comparison against SNR }
The BER performances of the  ADMM detector are compared with those of the traditional MIMO detectors like the MMSE detector under different propagation models.  Fig.~\ref{siso_admm} shows the BER performance of IRS-OTFS with all the cases under BPM. Case 1 is the one in which the RC matrix is optimized via STM and ADMM is used for signal detection, Case 1 is the proposed algorithm. At the BER of $10^{-5}$, the coding gain of Case 1 over all other cases is observed. The coding gains of Case 1 over Cases 2, 3, and 4 are around 1 dB, 5 dB, and 8 dB, respectively. Case 5 outperforms Case 1 since,  in Case 5, D-link is considered. The coding gain of Case 5 is approximately 2~dB over Case~1.\newline
Finally, BER is simulated for  BPM  considering imperfect cascaded CSI and D-link is not taken into consideration (Case~6). It is found that the coding gain of Case~6 is poor as compared to Case~1 and Case~2.

\vspace{-4mm}
 \begin{table}[!htbp]
     \centering
     \caption{Various Cases for BER Analysis.}
     \vspace{-2mm}
\begin{tabular}{ |c| c|c|c|c|c|c| }
\hline
Cases   & IRS RC  & Detection technique & D-link & CSI   \\ \hline 
Case 1  & STM & ADMM &Absent& Perfect \\ \hline
Case 2 & Random & ADMM & Absent & Perfect\\   \hline 
Case 3 & STM  & MMSE & Absent & Perfect\\

\hline
Case 4 & Random & MMSE & Absent & Perfect \\ \hline
Case 5 & STM & ADMM & Present & Perfect \\ \hline
Case 6 & STM & ADMM & Absent & Imperfect \\  \hline
\end{tabular}
  \label{tab6}
\end{table}

\pgfplotsset{every semilogy axis/.append style={
		line width=0.9 pt, tick style={line width=0.9pt}}, width=8cm,height=6.5cm, 
	legend style={font=\scriptsize},
	legend pos= south west}
\begin{figure}[htb]
\centering
\begin{minipage}{.5\textwidth}
  \centering
  \begin{tikzpicture}[new spy style]
			\begin{semilogyaxis}[xmin=0, xmax=20, ymin=1.4e-08, ymax=0.2,
				xlabel={\small SNR (dB)},
				ylabel={\small BER},
				grid=both,
				grid style={dotted},
				legend cell align=left,
    legend entries={Case 1, Case 3, Case 2, Case 4,Case 5,Case 6, OFDM IRS },
				cycle list name=mycolorlist,
				]
	 \node[text width=3cm] at (axis cs:13,10e-6){\scriptsize $M=32$ };
  \node[text width=3cm] at (axis cs:13,4.5e-6){\scriptsize $N=32$ };
  \node[text width=3cm] at (axis cs:13,2e-6){\scriptsize $K=16$ };
 \addplot [magenta,dashed,mark=oplus,thick,mark size=3pt, mark options={scale=1,solid}] table [x={x1}, y={y1}] {SISO_admm_k=4_comp1.txt};
\addplot [cyan,mark=o,thick,mark size=3pt] table [x={x1}, y={y2}] {SISO_admm_k=4_comp1.txt};
\addplot [green,mark=diamond,thick,mark size=3pt]  table [x={x1}, y={y3}] {SISO_admm_k=4_comp1.txt};
\addplot [blue,mark=pentagon,thick,mark size=3pt] table [x={x1}, y={y4}] {SISO_admm_k=4_comp1.txt};
\addplot [red,mark=square,thick,mark size=3pt]  table [x={x1}, y={y5}] {SISO_admm_k=4_comp1.txt};
\addplot [black,mark=triangle,thick,mark size=3pt]  table [x={x1}, y={y6}] {SISO_admm_k=4_comp1.txt};
\addplot [black,mark=star,thick,mark size=3pt]  table [x={x1}, y={y7}] {SISO_admm_k=4_comp1.txt};
               
			\end{semilogyaxis}
		\end{tikzpicture}		
		\vspace{-6mm}
		\caption{\footnotesize{ BER of SISO OTFS-IRS with ADMM Detector in BPM.}	}	
	\label{siso_admm}
\end{minipage}%
\begin{minipage}{.5\textwidth}
  \centering
 \begin{tikzpicture}[new spy style]
			\begin{semilogyaxis}[xmin=0, xmax=10, ymin=6e-7, ymax=0.5,
				xlabel={\small SNR (dB)},
				ylabel={\small BER},
				grid=both,
				grid style={dotted},
				legend cell align=left,
    legend entries={Case 1 with Beamforming, Case 2 with Beamforming,Case 3 with Beamforming,Case 4 with Beamforming, Case 1 without Beamforming, Case 2 without Beamforming },
				cycle list name=mycolorlist,
				]
\addplot [magenta,dashed,mark=oplus,thick,mark size=3pt, mark options={scale=1,solid}] table [x={x1}, y={y1}] {mimo_with_bmforming1.txt};
\addplot [blue,mark=diamond,thick,mark size=3pt] table [x={x1}, y={y2}] {mimo_with_bmforming1.txt};
\addplot [orange,mark=star,thick,mark size=3pt]  table [x={x1}, y={y3}] {mimo_with_bmforming1.txt};
\addplot [black,mark=pentagon,thick,mark size=3pt]  table [x={x1}, y={y4}] {mimo_with_bmforming1.txt};
\addplot [red,mark=square,thick,mark size=3pt]  table [x={x1}, y={y5}] {mimo_with_bmforming1.txt};
\addplot [cyan,mark=o,thick,mark size=3pt]  table [x={x1}, y={y6}] {mimo_with_bmforming1.txt};
\end{semilogyaxis}
		\end{tikzpicture}		
		\vspace{-10mm}
		\caption{\footnotesize{ BER of MIMO IRS OTFS-system in BPM.}}	
	\label{mimo_admm_bmforming}
\end{minipage}
\vspace{-8mm}
\end{figure}


\subsubsection{BER Comparison against SNR along with Beamforming}
 Fig. \ref{mimo_admm_bmforming} presents the BER performance for the ADMM detector under BPM. The BER performance is compared under two scenarios: with and without optimized beamforming matrix \textbf{W}. The optimization of \textbf{W} is done using CVX optimization tool as per Algorithm~1. 
  The comparison of BER performance of the  IRS-OTFS system using optimized $\mathbf{W}$ across different scenarios under the BPM channel is shown in Fig.~\ref{mimo_admm_bmforming}. The coding gains of Case~1 over Case~2, Case~3, and Case~4 are approximately 0.5 dB, 2 dB, and 4 dB, respectively at the BER of $10^{-3}$.
 
 The coding gain of beamformed IRS-OTFS  over the IRS-OTFS  without beamforming is observed at the BER of $10^{-3}$.  It is found that Case~1 and Case~2 with optimized \textbf{W} have a 2 dB and 3.5 dB coding gain, respectively,  over the case of without optimized \textbf{W}. It's evident that optimizing $\mathbf{W}$ leads to a noteworthy enhancement in BER performance across all cases.

\subsubsection{BER Comparison against Number of Iteration}
We plot the BER as a function of the number of iterations to show the convergence of the algorithm. To compare the convergence of the proposed algorithm (Case 1), we add curves for 'Case 2', 'Case 3', and 'Case 4'.  Fig. \ref{wrt_iteration} illustrates that when the convergence error is less than or equal to one percent, the proposed algorithm is capable of converging within 40 iterations.
\vspace{-6mm}
\pgfplotsset{every semilogy axis/.append style={
		line width=0.9 pt, tick style={line width=0.9pt}}, width=8cm,height=6.5cm, 
	legend style={font=\scriptsize},
	legend pos= south west}
\begin{figure}[htb]
\centering
\begin{minipage}{.5\textwidth}
  \centering
  	\begin{tikzpicture}[new spy style]
			\begin{semilogyaxis}[xmin=0, xmax=100, ymin=1.0e-05, ymax=1.0e-01,
				xlabel={\small  Number of Iteration},
				ylabel={\small BER},
				grid=both,
				grid style={dotted},
				legend cell align=left,
    legend entries={Case 2, Case 1, Case 3, Case 4 },
				cycle list name=mycolorlist,
				]
\addplot [magenta,dashed,mark=oplus,thick,mark size=3pt, mark options={scale=1,solid}] table [x={x1}, y={y1}] {otfsirs_admm_ite.txt};
\addplot [cyan,mark=o,thick,mark size=3pt]  table [x={x1}, y={y2}] 
               {otfsirs_admm_ite.txt};
\addplot [black,mark=square,thick,mark size=3pt]  table [x={x1}, y={y3}] {otfsirs_admm_ite.txt};
\addplot [blue,mark=diamond,thick,mark size=3pt] table [x={x1}, y={y4}] {otfsirs_admm_ite.txt};
\end{semilogyaxis}
\end{tikzpicture}		
\vspace{-10mm}
\caption{\footnotesize{ BER of  IRS-OTFS using ADMM detector w.r.t. iteration in BPM.}}		
	\label{wrt_iteration}
 \vspace{-4mm}
\end{minipage}%
\begin{minipage}{.5\textwidth}
  \centering
\begin{tikzpicture}[new spy style]
			\begin{semilogyaxis}[xmin=-40, xmax=10, ymin=1.0e-06, ymax=100,
				xlabel={\small  SNR (dB)},
				ylabel={\small MSE},
				grid=both,
				grid style={dotted},
				legend cell align=left,
    legend entries={ BPM $N_{r} =N_{t}=1$ , BPM $N_{r} =N_{t}=2$ , BPM $N_{r} =N_{t}=3$  , Int. Dopp. EVA $N_{r} =N_{t}=3$,Frac. Dopp. EVA $N_{r}= N_{t}=3$ , IRS-OFDM $N_{r} =N_{t}=2$, IRS-OFDM $N_{r} = N_{t}=3$  },
				cycle list name=mycolorlist,
				]
\addplot [magenta,dashed,mark=oplus,thick,mark size=3pt, mark options={scale=1,solid}] table [x={x1}, y={y1}] {mse_comp.txt};
\addplot [cyan,mark=o,thick,mark size=3pt]  table [x={x1}, y={y2}] 
               {mse_comp.txt};
\addplot [black,mark=square,thick,mark size=3pt]  table [x={x1}, y={y3}] {mse_comp.txt};
\addplot [blue,mark=diamond,thick,mark size=3pt] table [x={x1}, y={y4}] {mse_comp.txt};
\addplot [red,mark=star,thick,mark size=3pt] table [x={x1}, y={y5}] {mse_comp.txt};
\addplot [orange,mark=pentagon,thick,mark size=3pt] table [x={x1}, y={y6}] {mse_comp.txt};
\addplot [purple,mark=triangle,thick,mark size=3pt] table [x={x1}, y={y7}] {mse_comp.txt};
         \node[text width=3cm] at (axis cs:10,10){\scriptsize $L=16$ }; 
          \node[text width=3cm] at (axis cs:10,30){\scriptsize $K=16$ }; 
          
			\end{semilogyaxis}
		\end{tikzpicture}		
		\vspace{-10mm}
		\caption{\footnotesize{ MSE of the estimated cascaded channel of IRS-OTFS .}}		
	\label{mse1}
\vspace{-6mm}
\end{minipage}
\vspace{-4mm}
\end{figure}


\vspace{-4mm}
\subsection{Mean Square Error (MSE) Analysis}
The accuracy of the channel estimation is examined using the mean square error (MSE), which is given as \vspace{-3mm}
\begin{equation}\label{mse}
    {\text{MSE}}(\hat{\textbf{G}}) = \frac{1}{I}\sum_{i=1}^{I}\frac{\left\|\textbf{G}^{(i)}-\hat{\textbf{G}}^{(i)} \right\|^{2}_{F}}{\left\|\textbf{G}^{(i)}  \right\|^{2}_{F}},
    \vspace{-3mm}
\end{equation}
where $\hat{\textbf{G}}^{(i)}$ is the estimated channel of BS-IRS for the $i$-th iteration and $I$ denotes the total number of iterations. Following (\ref{mse}), the accuracy of the channel estimation for IRS-UE channel ( \textbf{D}) can be examined.
Fig.~\ref{mse1} shows the MSE comparison of the cascaded channel of IRS-OTFS for both the BPM as well as EVA channel model. 
In the EVA model, both the integer as well as fractional Doppler are considered and the integer Doppler outperforms 
 the fractional Doppler case.
Fig. \ref{mse1} shows the MSE comparison of the estimated cascaded channel for IRS-OTFS  and IRS-OFDM  with different numbers of transmit and receive antennas as well. It is found from simulation that IRS-OTFS  outperforms IRS-OFDM. The performance  improves as the numbers of transmit as well as receive antenna increases.

\vspace{-6mm}
\section{Conclusion}\label{sec6}
This paper  focused on three critical issues of an IRS-aided OTFS system: (1) joint optimization of beamforming and IRS phase shifts, (2) low-complexity signal detection and (3) channel estimation of cascaded path. The beamforming design problem in the proposed IRS-aided OTFS system, focusing on maximizing the sum rate to improve capacity under the constraints of transmitted power and IRS  phase shifts, is addressed. To tackle this challenge, an  AO algorithm is proposed.
In addition, an efficient signal detector and channel estimation algorithm for IRS-aided OTFS systems using the  ADMM approach and ALS approach are developed.
Simulation results have demonstrated that the proposed algorithm for joint beamforming matrix optimization (at BS) and RC optimization (at IRS) and the proposed signal detection method outperform the conventional algorithms of the OTFS system in terms of sum rate and BER. The impact of various system parameters, such as OTFS grid size, UE location, and phase settings, is thoroughly analyzed for the IRS-aided OTFS system to demonstrate the effectiveness of the proposed beamforming and signal detection method.

\vspace{-6mm}
\bibliographystyle{IEEEtran}
\footnotesize
\bibliography{IEEEabrv} 

\end{document}